\documentclass[useAMS,usenatbib]{mn2e}
\usepackage[dvips]{graphicx}
\usepackage{graphicx}
\usepackage{amsmath}
\usepackage{amsfonts}
\usepackage{amssymb}
\newcommand{\etal}{et al.}
\newcommand{\rjbch}[1]{{#1}}
\title{Water in HD~209458b's atmosphere from 3.6 $-$ 8 $\mu$m IRAC photometric observations in primary transit}
\author[Beaulieu, Kipping, Batista, Tinetti, Ribas et al., ]{J.P. Beaulieu$^{1,2,3}$, D.M. Kipping$^{2,3}$, V. Batista$^{1,3}$, G. Tinetti$^{2,3}$, I. Ribas$^{4}$, S. Carey$^{5}$, 
 \newauthor J.  A.  Noriega-Crespo$^{5}$,  C. A. Griffith$^{6}$, G. Campanella$^{2,7}$, S. Dong$^{8}$, J. Tennyson$^{2}$, 
\newauthor  R.J. Barber$^{2}$,  P. Deroo$^{9}$,  S.J. Fossey$^{2}$, D. Liang$^{10,11}$, M. R. Swain$^{9}$,  Y. Yung$^{11}$,
\newauthor N. Allard$^{12,1}$\\
$^{1}$Institut d'Astrophysique de Paris, CNRS, UMR7095, Universit\'e Paris VI, 98bis Boulevard Arago, PARIS, France\\
$^{2}$Department of Physics and Astronomy, University College London, Gower street, London WC1E 6BT, UK \\
$^{3}$HOLMES Collaboration\\
$^{4}$  Institut de Ciencies de l'Espai (CSIC-IEEC), Campus UAB, 08193 Bellaterra, Spain\\
$^{5}$   IPAC-Spitzer Science Center, California Institute of Technology, Pasadena, CA 91125, USA\\
$^{6}$ Department of Planetary Sciences, Lunar and Planetary Laboratory, The University of
Arizona, Tucson, AZ 85721-0092, USA \\
$^{7}$  Dipartimento di Fisica, Universit\`a di Roma ''La Sapienza'', Ple Aldo Moro 5, 00185 Rome, Italy\\
$^{8}$  Ohio State University, Colombus, Ohio, USA \\
$^{9}$  Jet Propulsion Laboratory, California Institute of Technology, 4800 Oak Grove Drive, Pasadena, California 91109-8099, USA\\
$^{10}$  Research Center for Environmental Changes, Academia Sinica, Taipei, Taiwan \\
$^{11}$ Division of Geological and Planetary Sciences, California Institute for Technology, Pasadena, CA 91125, USA \\
$^{12}$  GEPI, Observatoire de Paris, 77 Avenue Denfert Rochereau, 75014 PARIS, France\\
       }

\begin{document}

\date{Submitted on}

\pagerange{\pageref{firstpage}--\pageref{lastpage}} \pubyear{2008}

\maketitle

\label{firstpage}

\begin{abstract}

The hot Jupiter HD~209458b was observed during primary transit at 3.6, 4.5, 5.8 and 8.0 $\mu$m using the Infrared Array Camera (IRAC) on the Spitzer Space Telescope. We describe the procedures we adopted to correct for the systematic effects present in the IRAC data and the subsequent analysis. The lightcurves were fitted including limb darkening effects and fitted using Markov Chain Monte Carlo and prayer-bead Monte Carlo techniques, obtaining almost identical results. The final depth measurements obtained by a combined Markov Chain Monte Carlo fit are at 3.6 $\mu$m, $1.469 \pm 0.013$ \% and $1.448 \pm 0.013$ \%; at  4.5 $\mu$m, $1.478 \pm 0.017$ \% ; at   5.8 $\mu$m, $1.549 \pm 0.015$ \% and at 8.0 $\mu$m $1.535 \pm 0.011$ \%.  Our results clearly indicate the presence of water in the planetary atmosphere. 
Our broad band photometric measurements with IRAC prevent us from determining the additional presence of other other molecules
such as CO, $\mathrm{CO}_2$ and methane for which spectroscopy is needed. 
While water  vapour  with a mixing ratio of $10^{-4}-10^{-3}$ combined with thermal profiles retrieved from the day-side may provide a very good fit to our observations, this data set alone is unable to resolve completely the degeneracy between water abundance and atmospheric thermal profile.
\end{abstract}


\section{Introduction}

More than 420 exoplanets, i.e. planets orbiting a star \rjbch{other
than} our Sun, are now known thanks to indirect detection techniques
(Schneider, 2009). In the first decade after the initial discovery of a
hot Jupiter orbiting a solar like star in 1995 (Mayor and Queloz, 1995), the task was to find more and more
of these astronomical bodies. In recent years, attention has switched
from finding planets to characterising them. Among the variety of
exoplanets discovered, \rjbch{particular} attention \rjbch {is being}
devoted to those planets \rjbch{that} transit their parent star,
\rjbch {and} whose presence can \rjbch{therefore} be detected by
\rjbch{a} reduction in the brightness  of the central star as the planet passes in front of
it. Sixty-nine of the 420+ currently identified exoplanets are
transiting planets, and for these objects planetary and orbital
parameters such as radius, eccentricity, inclination, mass (given by
radial velocity combined measurements) are known, allowing first
order characterisation on the bulk composition and temperature. 
In particular, it is possible to exploit the wavelength dependence of this
extinction to identify key chemical components in the planetary
atmosphere (Seager and Sasselov, 2000; Brown, 2001), \rjbch{which} permits
enormous possibilities for exoplanet characterisation.

The extrasolar planet HD~209458b orbits a main sequence G star at
0.046 AU (period 3.52 days). It \rjbch{was} the first
\rjbch{exoplanet} for which repeated transits across the stellar disk
were observed ($\sim$ 1.5\% absorption; Charbonneau et al., 2000).
\rjbch{Using} radial velocity measurements (Mazeh et al., 2000), the
planet's mass and radius \rjbch{were able to be} determined ($M_{p}
\sim 0.69 \, M_{Jup}$, $R_{p} \sim 1.4 \, R_{Jup}$), confirming the
planet is a gas giant with one of the lowest densities \rjbch{so far}
discovered. \rjbch{Consequently it must possess} a highly extended atmosphere
making it one of the optimum candidates \rjbch{for observation using} primary
transit techniques, and it was indeed the first exo-atmosphere probed
successfully \rjbch{using this method} in the visible (Charbonneau et al.,
2002) and then in the infrared (Richardson et al., 2006).

Following the work on HD~189733b, where the first detections of water
vapour (Tinetti et al., 2007b; Beaulieu et al., 2008) and methane
(Swain, Vasisht \& Tinetti 2008) have been achieved, we were
awarded 20 hours Director's Discretionary Time on Spitzer (PI
Tinetti, WETWORLD, PID 461) to probe the atmosphere of HD~209458b in
primary transit in the four IRAC bands at 3.6, 4.5, 5.8 and 8 $\mu$m (channels 1 to 4 respectively).
Water vapour was proposed to be present in the atmosphere of HD~209458b 
by Barman (2007), to fit the data recorded by Hubble-STIS in
the visible (Knutson et al., 2007).  Also, water vapour  combined with a thermal profile increasing with altitude
was a reasonable explanation to fit the secondary transit photometric data
observed in the mid-IR (Deming et al., 2005; Knutson et al., 2007;
Burrows et al, 2007). Our understanding of the thermal profile and
composition has improved thanks to more recent secondary transit
spectroscopic data in the near and mid-IR, indicative of the additional 
presence of methane and carbon dioxide in the atmosphere of 
HD~209458b (Swain et al., 2009b), confirmed by Madhusudhan N.  \& Seager S. (2010).
 
 Transmission and emission spectra probe different regions of a
hot-Jupiter atmosphere, both longitudinally and vertically (Tinetti \&
Beaulieu, 2008).  In particular, \rjbch{the mid-infrared} primary
transit observations described here allow us to probe the terminator region of
HD~209458b between the bar and millibar level.

\section{Observations and data reduction}

\begin{figure*}
\begin{center}
\includegraphics[angle=0,width=9. cm]{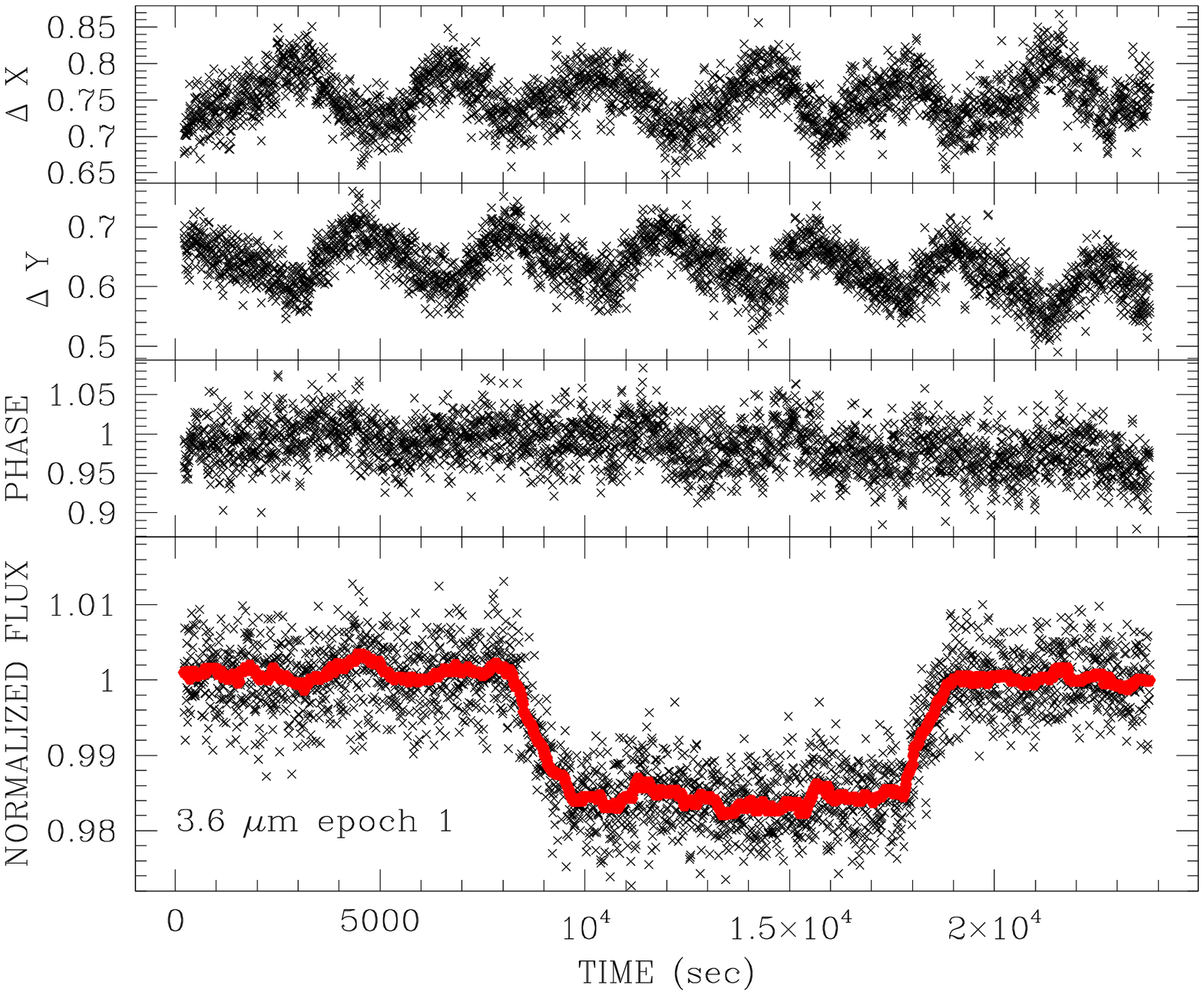}\includegraphics[angle=0,width=9. cm]{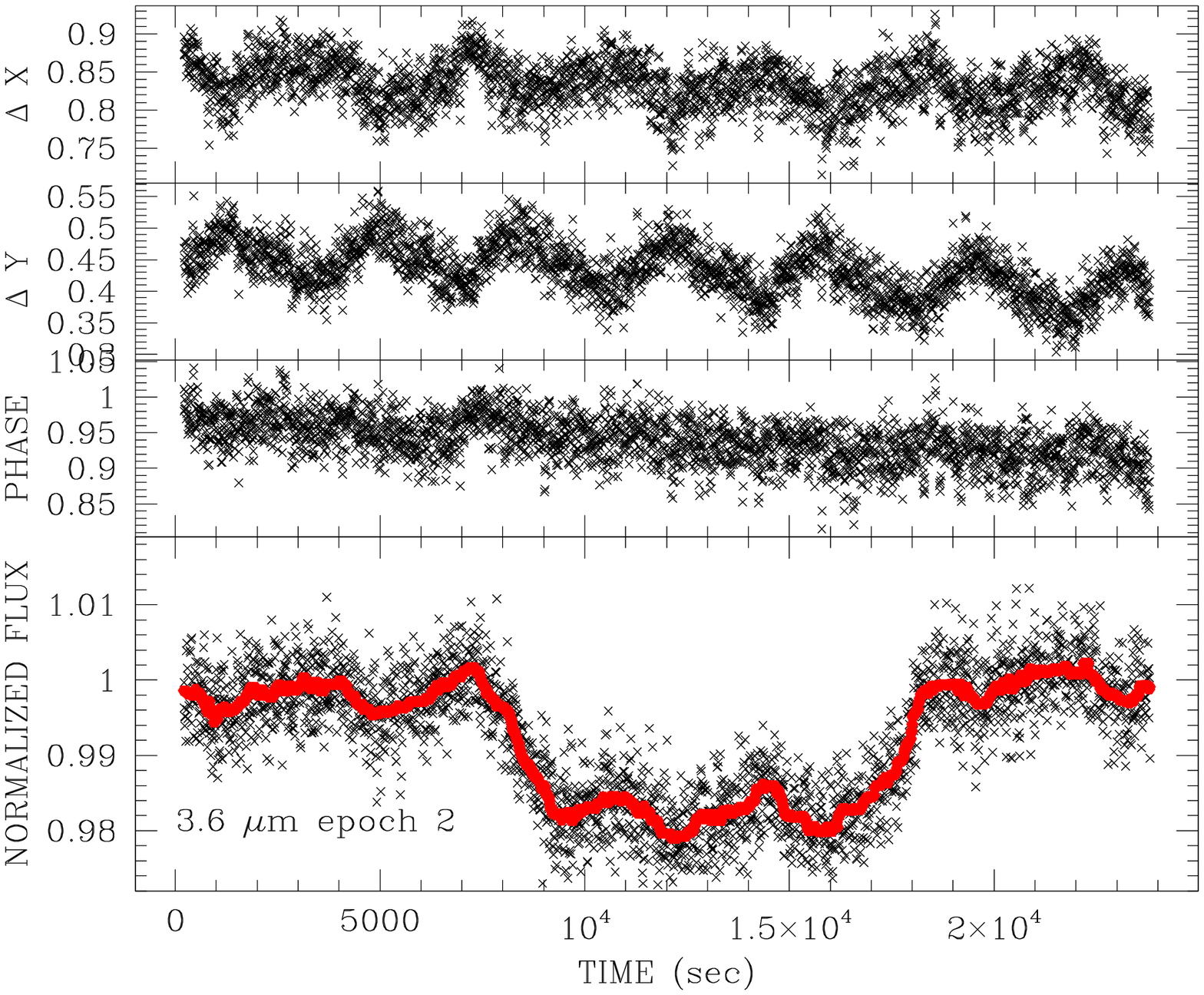}
\includegraphics[angle=0,width=9. cm]{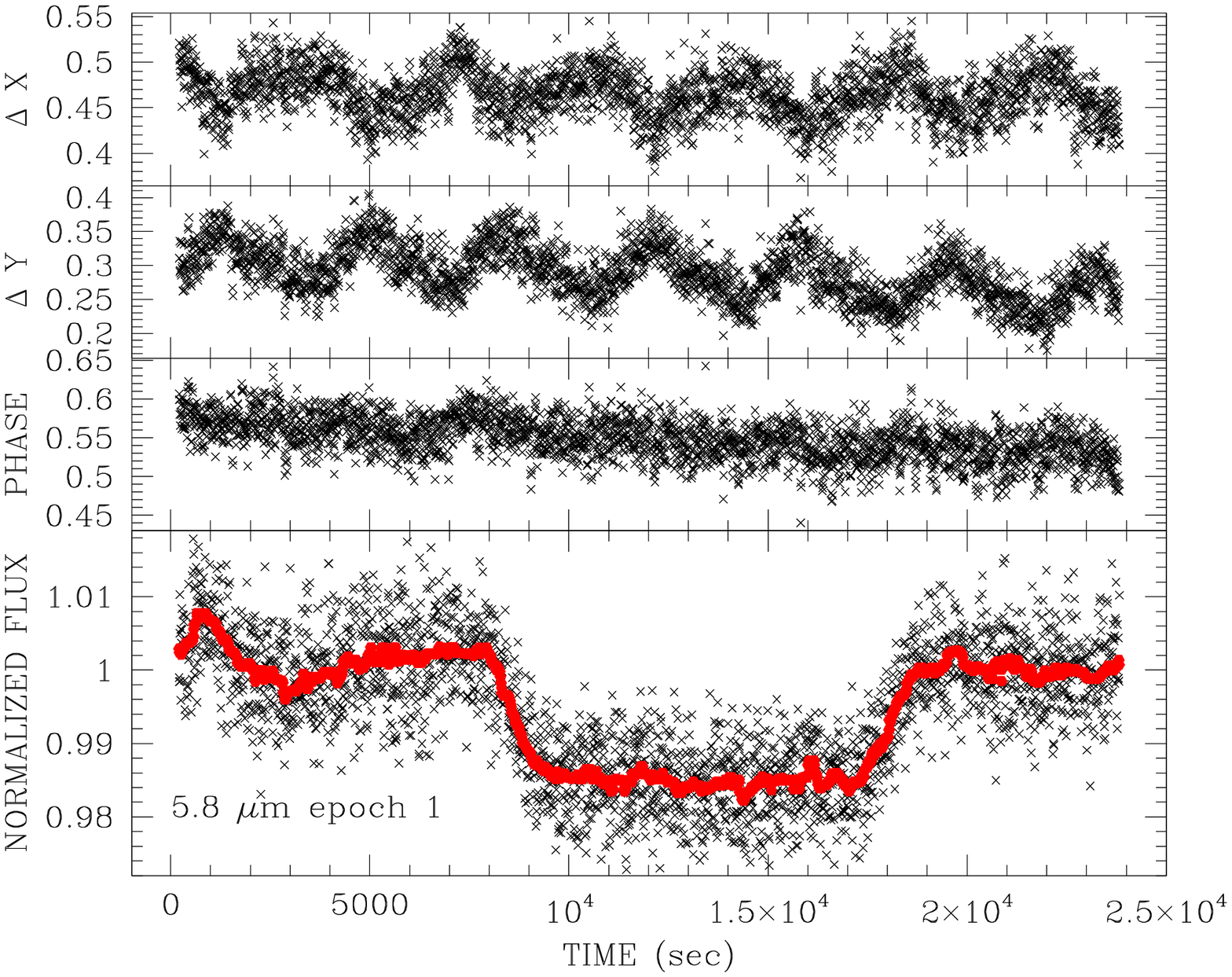}\includegraphics[angle=0,width=9. cm]{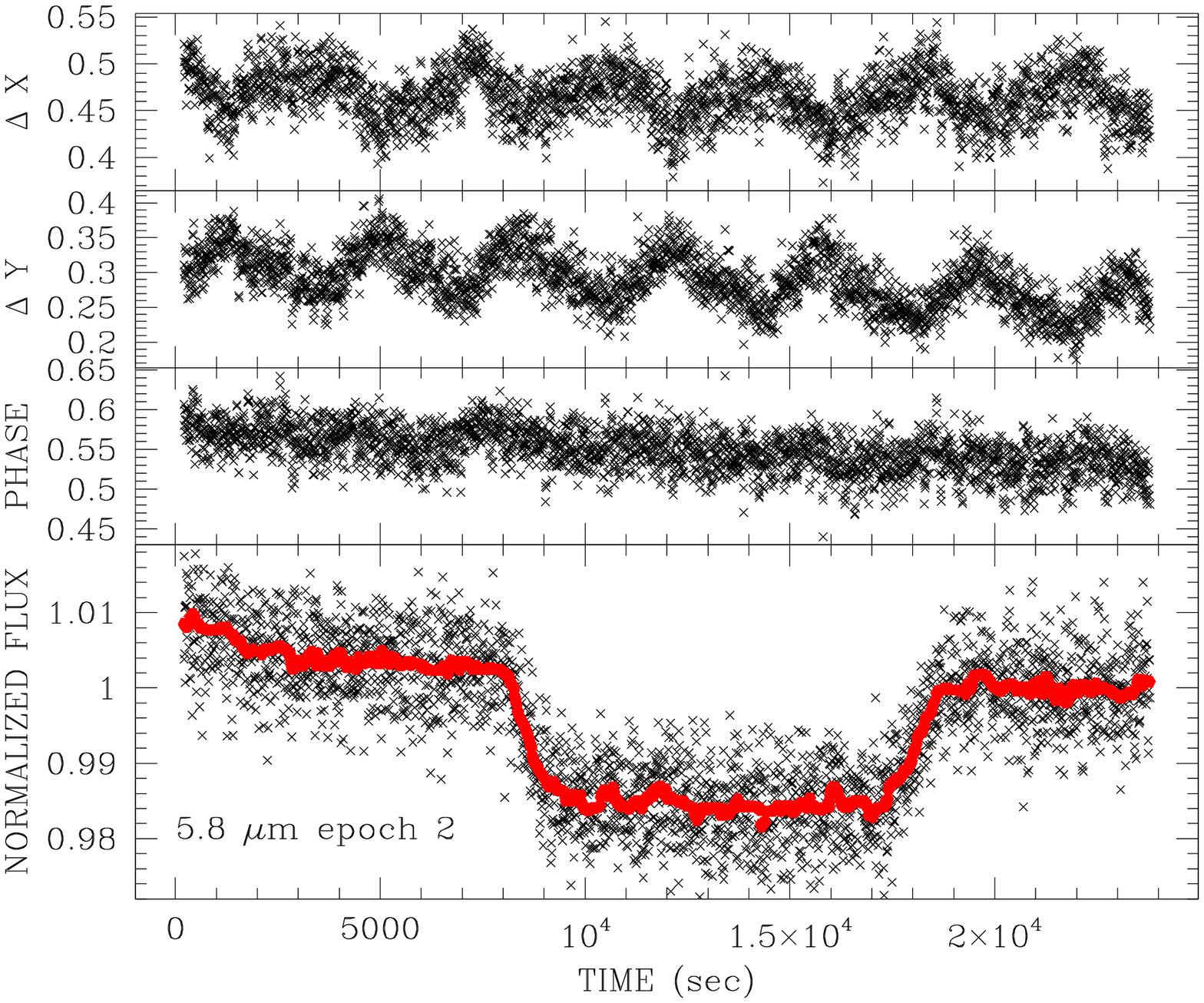}
\includegraphics[angle=0,width=9. cm]{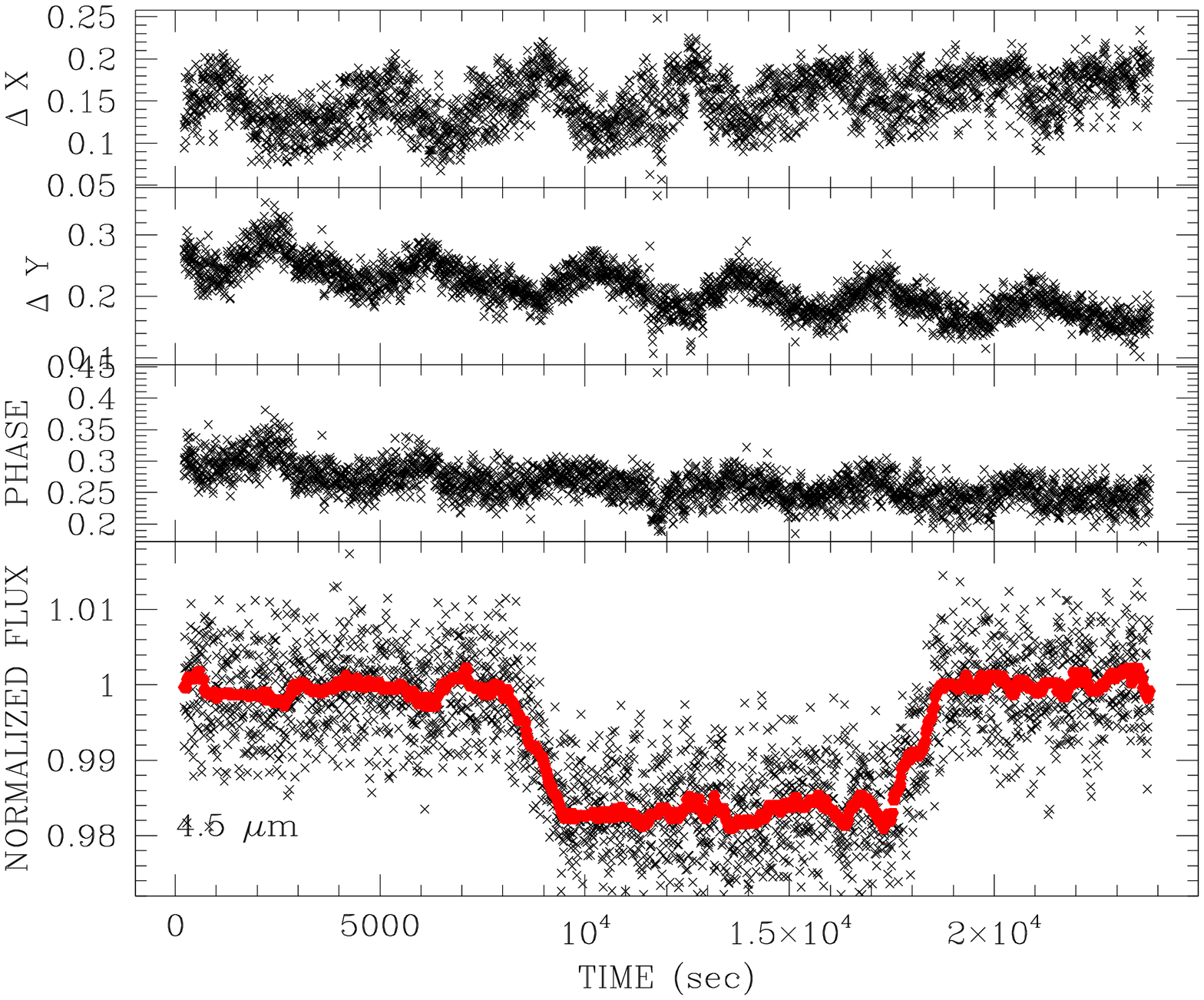}\includegraphics[angle=0,width=9. cm]{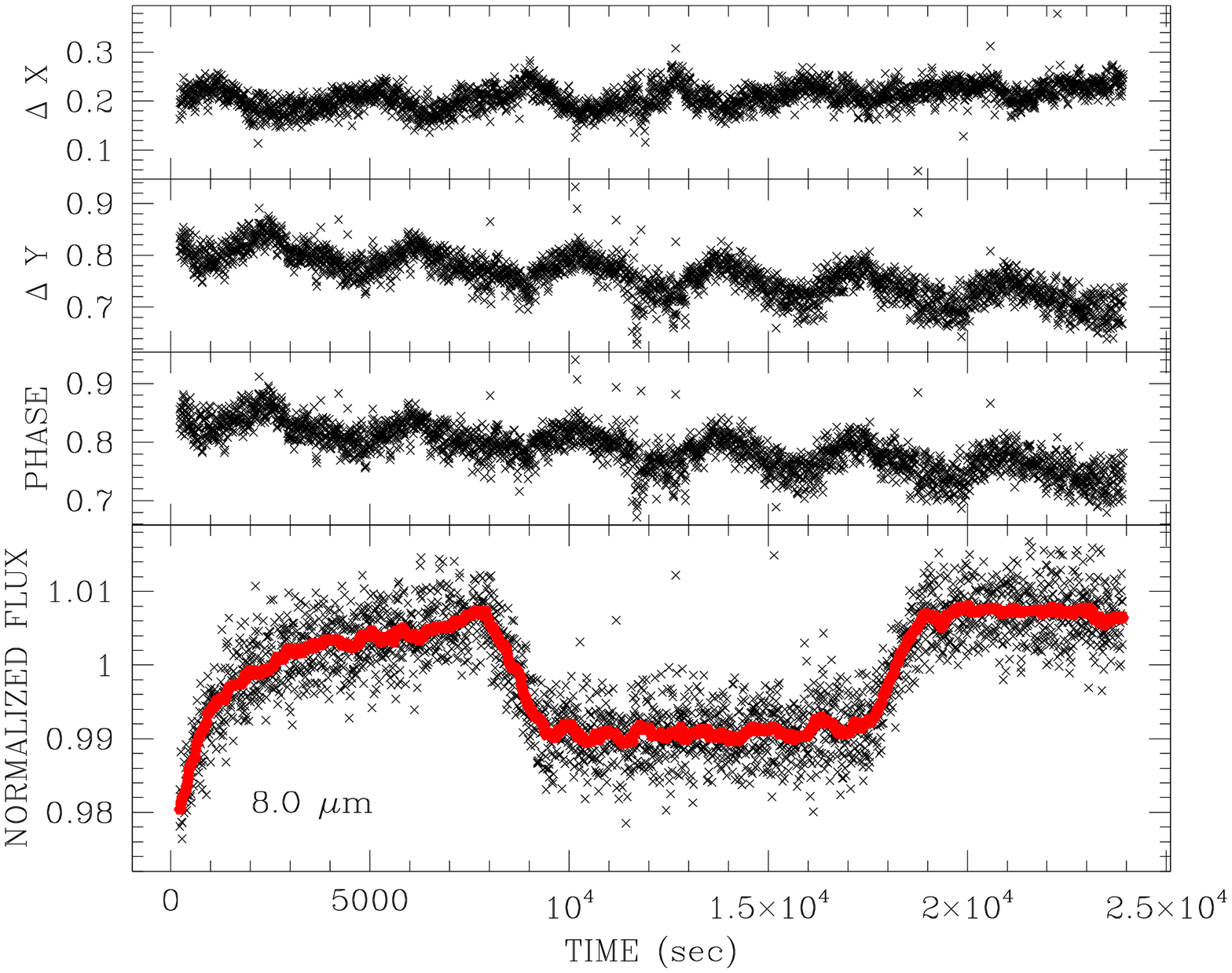}
\caption{\emph{Raw photometric data for 3.6 $\mu$m 
(epoch 1 and 2), 5.8 $\mu$m (epoch 1 and 2), 4.5 $\mu$m  and 8 $\mu$m obtained
with IRAC. Each sub-panel has the same structure showing
from top to bottom: the variation of the centroid position in X,in Y,
and lastly the distance of the centroid from the lower left corner in the pixel
(called the pixel phase, that can also be seen as the pointing error temporal amplitude). 
The lowest panel of each plot is the primary
transit, and over-plotted the 50-point median-stack smoothing. They
provide a synoptic view of the systematic trends present in IRAC primary transit
data.  }} \label{fig:figraw}
\end{center}
\end{figure*}

\subsection{Planning the observations}

Three HD~209458 primary transits \rjbch{were} observed with the IRAC
camera on board the \rjbch{Spitzer Space Telescope. Channels} 1 and 3
(3.6 and 5.8 $\mu$m) were \rjbch{observed} at two epochs, on December 30,
2007 and July 18, 2008, and \rjbch{data were obtained using channels} 2
and 4 (4.5 and 8 $\mu$m) on July 20, 2008.  Since HD~209458 is a G0V
star with a 2MASS Ks magnitude of 6.3, the IRAC predicted fluxes are
878, 556, 351 and 189 mJy in channels 1-4, respectively.  For \rjbch{our}
observations we required extremely high signal-to-noise as 
the
modelled contribution to the absorption due to H$_2$O  was
predicted to be a few  times $10^{-4}$ of the stellar flux.

As \rjbch{with other Spitzer observations of transiting planets, it is
  necessary to observe the target continuously without dithering, in
  order} to be able to quantify optimally the systematic effects detailed below. 

- Flat-fielding errors are an important issue; observations at
different positions on the array \rjbch{effectuate systematic
  scatter in the photometric data that can potentially swamp the
  signal that} we are looking for.

- The amount of light detected in \rjbch{channels} 1 and 2 shows
variability that depends on the relative position of the source with
respect to the pixel centre (labelled the pixel phase effect). The time
scale of this variation is of the order of \rjbch{50} minutes. These
effects are well known and documented in the IRAC Data Handbook and
also discussed by Morales-Calderon (2006), Beaulieu et al.,
(2008), Knutson et al., (2008), Agol et al., (2008). To first order, \rjbch{these are able to be corrected for
  using} the prescription of Morales-Calderon (2006). 
\rjbch{However, as the effects are variable across the array, ultimately they have to be} estimated from the data themselves.

- In \rjbch{channels} 3 and 4 there are only minor pixel phase effects,
but a variation of the response of the pixels to a long period of
illumination and latent build-up \rjbch{effect impinge on the} 5.8 and 8.0 $\mu$m
observations, respectively.

- We obtained a slightly longer
  `pre-transit' data set, in order to allow the satellite settle in a
  `repeatable' jitter pattern and a shorter post-transit data set. The
  time scale of the pixel-phase effect being of the order of \rjbch{50}
  minutes, we chose 120 min of pre- and 80 min of post-transit data
  baseline.

\rjbch{It is important to note} that the $\sim 184$ \rjbch{minute}
transit of HD~209458b means that \rjbch{our data contain} three full
cycles of the pixel phase variation in the transit itself,
giving an excellent opportunity to have a full control on the behaviour
of the systematic effects by evaluating them both in and outside the
transit.

\rjbch{Our observations employed the} IRAC 0.4/2 second stellar
photometry mode.  Using the regular Astronomical Observation Templates
(AOTs), a total of two transits per field-of-view was required to
achieve the desired sensitivity at 4.5 and 5.8 $\mu$m (the arrays with
the limiting sensitivity).  Unfortunately, the AOTs as designed were not
the most efficient way to perform this observation.  Each stellar mode
frame effectively \rjbch{incurs 8 seconds of overheads} due to data
transfer from the instrument to the spacecraft.  As \rjbch{our}
observations only \rjbch{required} the data in the field-of-view with
the star, \rjbch{it was possible to} save both data volume \rjbch{by
collecting data in} only two channels and with a cadence of 4
seconds. \rjbch{Consequently, we designed a} special engineering
template (Instrument Engineering Request; IER) to optimise the
observations. IERs have been used successfully in other planet transit
experiments (Charbonneau et al.  2005), \rjbch{and they can typically
double the efficiency, and our IER enabled us to reduce the total
required observing time for all four channels to} only 13.4 hours.

\subsection{Data reduction and flux measurements}
We used the flat-fielded, cosmic-ray-corrected and flux calibrated
data files provided by the Spitzer pipeline. Each channel has been
treated separately. We measured the flux of the target on each image
using the version 2.5.0 of the SExtractor package (Bertin \& Arnouts, 1996), with a
standard set of parameters for Spitzer (Infrared Array Camera Data
Handbook, 2006).  The centroid determination \rjbch{was} achieved with PSF
fitting. We performed both aperture photometry, and PSF fitting
photometry. In Fig.1, for all the six observed transits, we give the
raw magnitude measurements (normalized using the post-transit observation), 
the variation of the centroid in $X$ and
$Y$ axis, and the distance of the centroid \rjbch{from} the lower left
corner of the pixel (the pixel phase).  A quick inspection shows that
all observations \rjbch{contain} correlated noise of different nature, as
expected when using the IRAC camera. We \rjbch{discuss this
  phenomenon, and how we corrected for it, channel by channel, in the
  next section}.

\section{Estimation and attenuation of correlated noise}
\subsection{Correcting the pixel phase effects}

It has been well-established that the IRAC channels exhibit pixel phase effects due to a combination of non-uniform response function within each pixel and very small pointing variations (Morales-Calder\'on et al., 2006, Beaulieu et al., 2008). These effects are most prominent within the 3.6 $\mu$m and 4.5 $\mu$m photometry and to a lesser degree in the other two channels.  We note that previous studies have not corrected for possible pixel phase effects at 5.8$\mu$m or 8$\mu$m, but in this work we evaluate the effectiveness of implementing it in all channels.

Pixel-phase information is retrieved by using \rjbch{SExtractor's} PSF fitting
to obtain estimates of the $X$ and $Y$ pixel-phase for each exposure.
In contrast, the flux for each exposure is obtained through aperture
photometry since this offers substantially larger signal-to-noise
compared to the PSF flux estimates.

A typical procedure is to directly correlate the $X$ and $Y$ phases to
the out-of-transit fluxes to some kind of 4 or 5 parameter fit (Morales-Calder\`on et al., 2006, Beaulieu et al., 2008, Knutson et al. 2008) and in this work we will adopt a similar approach.  We note that the PSF-fitted estimates of $X$ and $Y$ exhibit significant scatter at the same level as the amplitude of the periodic variations in each.  This scatter is caused predominantly by photon-noise slightly distorting the PSF shape in a random manner and thus causing the fitting algorithm to deviate from the true value.  The pixel phase effect is physically induced by the spacecraft motion and so we only wish to correlate to this property, as opposed to the random photon-noise induced scatter of $X$ and $Y$.  In order to do this, we fit a smooth function through the pixel-variations themselves before attempting to correlate to the out-of-transit flux.

Our analysis of the $X$ and $Y$ phases reveals a dominant $\sim$1 hour period sinusoidal-like
variation in $X$ and $Y$, characteristic of small elliptical motion in
Spitzer's pointing, with a more complex time trend overlaid.  For each
channel, we apply a non-linear regression of a sinusoidal wave to the phases, in order to determine the best-fit dominant period, $P_{phase}$ (typically close to one hour).  We then calculate the median of the phases from the
$i^{\mathrm{th}}$ data point to the $j^{\mathrm{th}}$, where $t_j = t_i + P_{phase}$, (where $t_k$ is the time-stamp of the $k^{\mathrm{th}}$ exposure) and repeat from $i=1$ up to the end of the data list.  This moving-window-function essentially purges the dominant period from the phases and thus allows us to obtain a robust determination of the second-order phase variations, which may then be fitted for using a polynomial, of orders varying from 2 to 4 depending on the degree of curvature in the resultant phase trends.

We have now calculated the function which describes the pixel phase variation of $X$ and $Y$ with respect to time, as induced by spacecraft motion.  This function is then correlated to the actual out-of-transit photometry to find a fit to the function $a+b X(t)+c Y(t) + d [X(t)]^2 + e [Y(t)]^2$.  We find including an additional cross-term does not further improve the pixel-phase-effect attenuation.

For 3.6$\mu$m (epochs 1 and 2) and 4.5$\mu$m, we \rjbch{removed} pixel-phase effects of r.m.s. amplitude 0.49, 1.51 and 0.57 mmag respectively, over the standard 8.4 second cadence.  The second epoch at 3.6$\mu$m, is particularly polluted by pixel phase response, possibly due to a large inhomogeneity in response close to the PSF centroid position (pixel 131,128 of the detector).  Repeating the process for the remaining channels (after the other systematic effects were removed first, see next sections for details), we are able to remove 0.29 and 0.24 mmag for 5.8$\mu$m (epoch 2) and 8$\mu$m respectively.  Thus the pixel-phase induced variations are half of the minimum variations founds at 3.6$\mu$m and 4.5$\mu$m.  
%We do not consider 5.8$\mu$m, epoch 2, due to the extensive systematic problems already effecting this data set. See also Table 1.

\begin{table*}
\caption{\emph{Noise properties and effects of pixel-phase effect attenuation on each IRAC channel. }} % title of Table
\begin{tabular}{c c c c c c c} % centred columns (5 columns)
\hline\hline
& 3.6  $\mu$m  (epoch 1) & 3.6  $\mu$m (epoch 2)  & 4.5  $\mu$m  & 8.0  $\mu$m  & 5.8  $\mu$m  (epoch 2) & 5.8  $\mu$m  (epoch 1)\\ 
\hline
\emph{Before correction} \\
\hline 
Baseline r.m.s./mmag   & 3.56005 & 3.8848  & 4.93071 & 3.26671 & 4.31389 & 4.82004 \\
in \% above photon noise & 75.5011 & 92.3528 & 119.734 & 79.3238 & 57.5393 & 76.1831 \\
\hline
\emph{After correction} \\
\hline
Baseline r.m.s./mmag   & 3.52621  & 3.57796  & 4.8974     & 3.25821 & 4.3042 & 4.7886 \\
in \% above photon noise & 73.8329   & 77.1596 & 118.249   & 78.8573 & 57.1856 & 75.0341 \\
Noise removed/mmag  & 0.489685 & 1.51325 & 0.572162 & 0.235493 & 0.288931 & 0.549594 \\
\hline\hline				
\end{tabular}
\label{table:pixel} % is used to refer this table in the text
\end{table*}

\subsection{Correcting systematic trends at 5.8 $\mu$m}
\begin{figure}
\begin{center}
\includegraphics[angle=0,width=9. cm,]{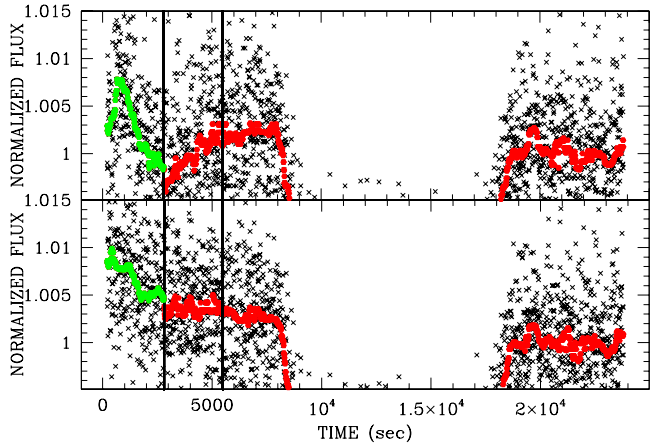}
\caption{Zoom on the IRAC 5.8 $\mu$m observations to show the systematic trends that are present. First and second epoch, in the upper and lower panels respectively. We show the data  and the  50-point median-stack smoothing. For the second epoch notice the change of behaviour around 2800 seconds, indicated by the vertical line. Note that the behavior after 2800 sec is different between the two epochs.} \label{fig:figcomp}
\end{center}
\end{figure}

In the exoplanet community, at least two different methods have been proposed to correct for the systematic effects observed at 5.8$\mu$m, characterized by a large change in flux near the commencement of the observations.  One frequently-adopted proceedure adopted is to discard the first $\sim $30 min of observations (Knutson et al., 2008, Charbonneau et al., 2008) and then de-trend the remaining data.  For example, in the case of HD 189733b primary transit observations, Beaulieu et al. (2008) removed the first 20 min, and then applied a linear correction.  Another method proposed by D\'esert et al. (2009) involves not excluding these first 20 minutes but attempt to correct the data using a logarithmic parameterisation (see sec. 4.4): $a +b t + c \log (t-t_0 ) + d (\log(t-t_0 ))^2$. Employing different corrective procedures will undoubtedly yield significantly different transit parameters and so we must carefully consider the effect of each proposed correction.

The most intuitive starting point is a visual inspection of the flux time series for our two measurements at 5.8$\mu$m.  In figure \ref{fig:figcomp}, we exclude the transit event and show the behaviour of the out-of-transit flux (the baseline) with an overlaid 50-point median-smoothing as a visual guide.  The first epoch exhibits a clear discontinuity between the photometry in the region $0 \leq t \lesssim 5500$ seconds and the subsequent data.  The behaviour of this initial photometry does not match a `linear drift, a `ramp' style-effect or any commonly employed analytic form.  The origin of the observed behaviour is unclear and is present in many different trial aperture sizes, between 2.5 to 20 pixels radius, suggesting an instrumental effect located either very close to the centroid position or globally across the detector array.

Repeating the visual inspection for the second epoch, we observe a less pronounced version of this behaviour in the region $0 \leq t \lesssim 2800$ seconds (note that this behaviour is not seen in any other channels).  However, the effect is ostensibly sufficiently small that we cannot claim it is the same behaviour from a visual inspection of the time series alone.  Therefore, we require a more in-depth analysis to provide a conclusion as to whether the systematic behaviours in epoch 1 and epoch 2 are the same.  In order to understand what kind of analysis this should be, we need to explicity qualify the question we are trying to answer.

The difference between the truncation + linear trend versus the logarithmic correction can be summarized by one key point: the former proposes that the initial data is incoherent with the latter data and cannot be characterized by a smooth analytic function.  The latter works under the hypothesis that the entire time series is following one single smooth analytic description.  We therefore wish to understand whether the properties of a smooth analytic function are consistent with the properties of the observed time series.  This is the critical question which we must answer.

One key property of the smooth analytic, logarithmic function proposed by D\'esert et al. (2009), is that the differential of the function with respect time provides another smooth analytic function.  In contrast, the truncation + linear trend hypothesis postulates that since the initial data exhibits discontinuous behaviour, then the differential of this must also be discontinuous.  So taking the differential of the time series will clearly resolve which hypothesis has the most supporting evidence.

To achieve this goal, we first extract the uncorrected out-of-transit fluxes only and remove outliers for both epoch 2 and epoch 1 using a median absolute deviation (MAD) analysis.  We then create a moving 150-point window, in which we calculate the local gradient at each point.  We do this by subtracting the median of the time stamps from each time stamp within a given window (to move the pivot along) and then performing a weighted linear regression.  We define the weights as the square of the reciprocal of each flux measurement.  In addition to the HD 209458b data presented here, we perform the same process on the HD 189733b 5.8$\mu$m data (used in Tinetti et al., 2007, Beaulieu et al., 2008, D\'esert et al. 2009) for comparison giving us three data sets.  Errors for each gradient stamp are computed using the weighted linear regression algorithm.

\begin{figure}
\begin{center}
\includegraphics[angle=0,width=9. cm,]{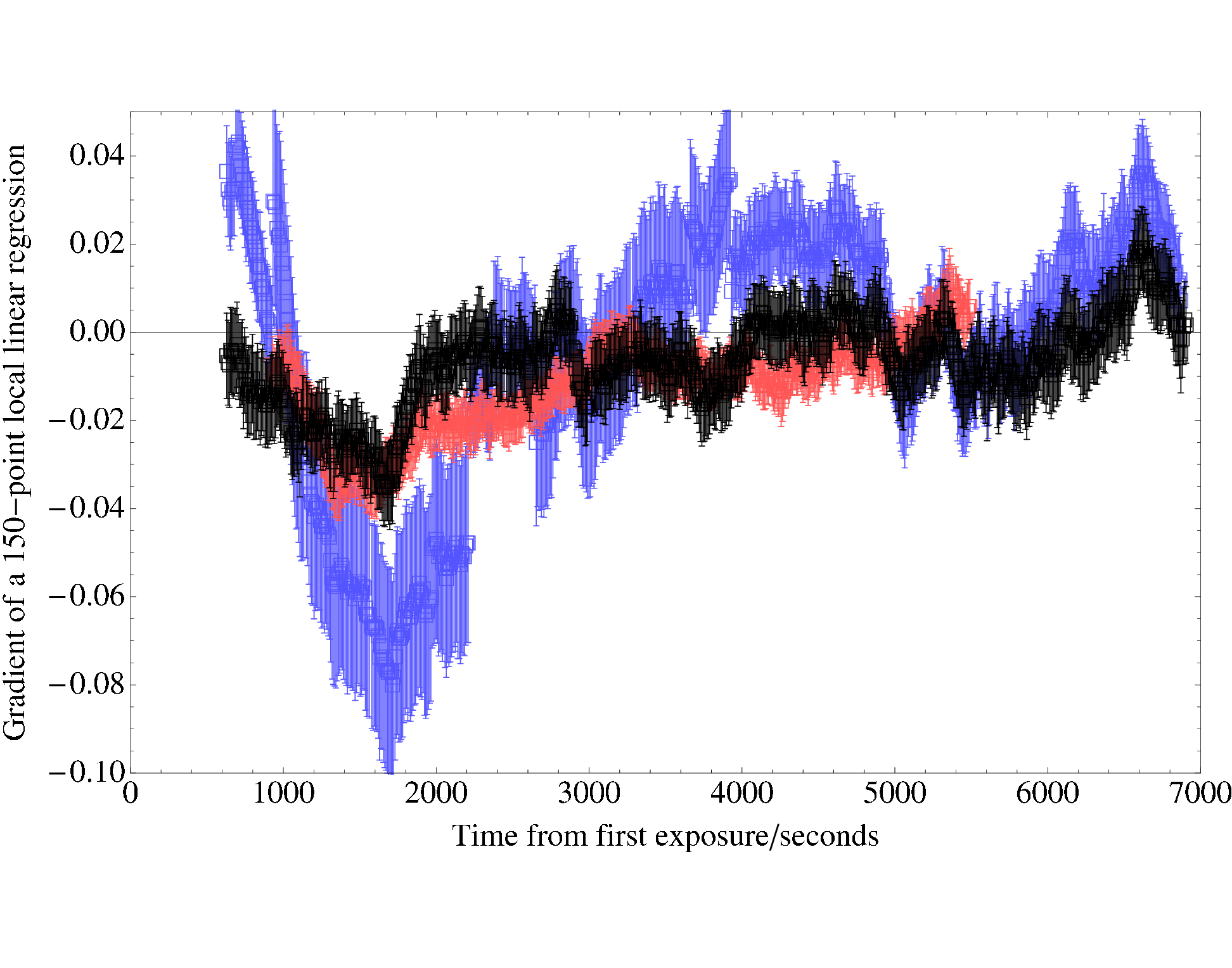}
\caption{Local gradient of each time stamp from the raw flux measurements obtained with IRAC at 5.8 $\mu$m 
for the two epochs of HD 209458b and HD 189733b. Note that the three exhibit similar behavior for the first 2000 sec.
The first epoch for HD 209458b has a larger amplitude of systematics, but the second epoch of HD 209458b and 
the observation of HD 189733b have remarkably similar behaviours.} \label{fig:figXY}
\end{center}
\end{figure}

In figure \ref{fig:figXY}, we see all three local gradients plotted together.  Ostensibly, there seems to be strong correlations between the three measurements, despite one of them being for a completely different star.  In particular, there is a strong dip at around 2000 seconds after the first exposure in all three observations.  In figure \label{fig:figgradient}, we plot just the epoch 2 of HD 209458 and also overlay the local gradients obtained from a logarithmic fit of the baseline (equivalent to the first differential of this function with respect to time).  It is clear that the logarithmic fit cannot explain the strong negative peak observed in the gradients data.  Furthermore, the clear presence of discontinuous behaviour in the local gradients supports the hypothesis that no continuous analytic function can correct this behaviour.

\begin{figure}
\begin{center}
\includegraphics[angle=0,width=8.5 cm]{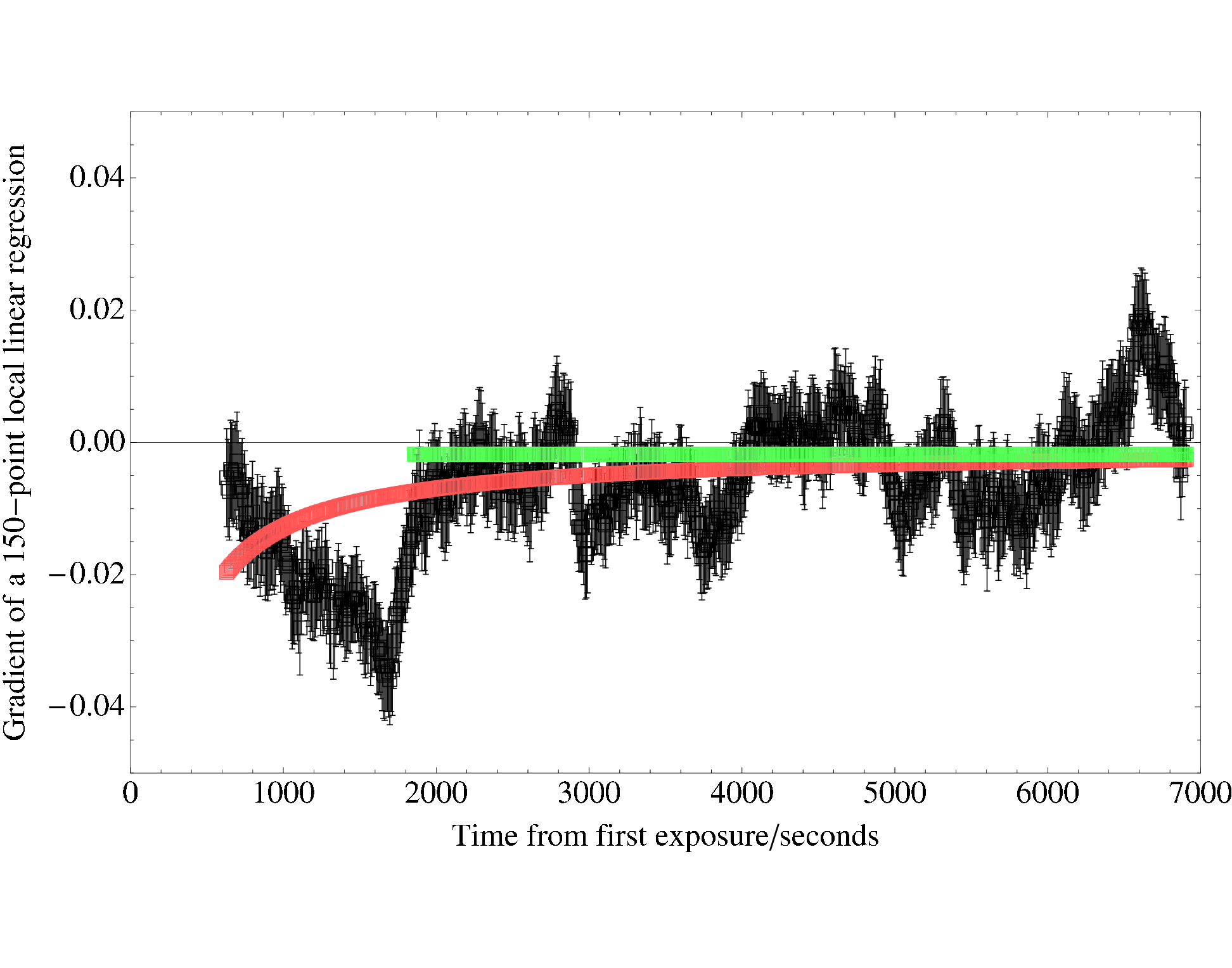}
\caption{ Local gradient of each time stamp from the raw flux measurements obtained with IRAC 5.8 $\mu$m (epoch 2) , computed using a linear regression of a moving 150-point window.  Black indicates the observed local gradients, which differ greatly from those obtained using a logarithmic fit of the photometry (red). Notice the 1600 seconds peak/discontinuity.} \label{fig:figgradient}
\end{center}
\end{figure}

Although the three measurements appear correlated, we may quantify these correlations.  Comparing any two channels, we define one as the reference data and one as the comparison data.  We first ensure the minimum to maximum time stamps of both sets are the same by clipping the longer set appropriately. We then perform a linear interpolation of both the gradient measurements and the uncertaintities, for the comparison data.  This allows us to accurately estimate the gradient values at like-for-like time stamps.  Regenerating the comparison gradients data using the interpolation function, we evaluate the correlation between the two using Pearson's correlation coefficient.  We repeat the same process for randomly generated data with the same uncertainities and array length as the original.  This is repeated 100,000 times in order to estimate the expected correlations from random noise.

Although they have been taken more than 7 months apart, we find epoch 1 and 2 for HD 209458 have correlation $\mathrm{Corr}(\mathrm{209458˜epoch˜2},\mathrm{209458˜epoch˜1}) = 0.678$.  The $10^5$ randomly generated noise values yield $0.001 \pm 0.037$.  This makes the correlation significant at the 18.4-$\sigma$ level.  Repeating the exercise for epoch 2 of HD 209458 and for  the observations of HD 189733b (taken 3 years apart), we have $\mathrm{Corr}(\mathrm{209458˜epoch˜2},\mathrm{189733}) = 0.658$.  The randomly generated noise yields $0.002 \pm 0.043$, making the observed correlation significant at the 15.2-$\sigma$ level.  In conclusion, the correlations in the local gradient plots are highly significant even for observations separated by years on different stars.  We therefore conclude the observed behaviour must be instrumental effects for 5.8$\mu$m detector array itself.

The largest feature is that of the `negative spike' at around 2000 seconds.  After this, all 3 observations exhibit variations consistent with that of a singular constant value i.e. a linear fit.  The reduced $\chi^2$ of these three channels may be computed both for all data and for those data after the negative spike.  We find the values always decrease by excluding the negative spike; quantitatively we have respective changes of $1.18 \rightarrow 0.58$ for HD 209458 epoch 2, $2.99 \rightarrow 1.24$ for HD 209458 epoch 1 and $2.66 \rightarrow 1.51$ for  HD 189733. Therefore we can see that the instrumental systematic effects of 5.8$\mu$m can be split into two regimes, pre and post spike.  The pre-spike data exhibits discontinuous behaviour compared to the latter data and cannot be characterized by a smooth continuous function.  The post-data conforms to a linear fit.

We therefore conclude that an analysis of the differential of the time series supports the hypothesis that the 5.8$\mu$m correction should be to remove the discontinuous data before the gradient spike and then use a linear fit for the remainder. It would therefore seem that at 5.8$\mu$m  the detector requires a certain amount of time to settle into a stable regime, as indicated also in earlier studies (Beaulieu et al., 2008, Knutson et al., 2008, Charbonneau et al., 2008).

Despite the evidence from this gradients analysis, we may conceive of several other possible tests to be certain that the logarithmic correction not favoured by the data.  Using the lightcurve fitting code described in \S4.3, we fitted two possible systematic correction lightcurve: 1) a truncation of the first 2800 seconds, followed by a linear fit to the remaining baseline data (previous examples Knutson et al. 2007; Harrington et al. 2007; Beaulieu et al. 2007) 2) a logarithmic fit to all baseline data (previous example D\'esert et al. 2009).  We select several properties to compare these two possible corrections:

\begin{enumerate}
\item Adopting a baseline between $2814\leq t \leq 7543$ seconds (i.e. after the discontinuous behaviour) and $19605 \leq t \leq 23974$ seconds, constituting 1082 data points, we compute the $\chi^2$ for both the linear and the logarithmic fit.  Despite using two extra free parameters, the logarithmic produces a larger $\chi^2=1358.1$ compared to a linear fit with $\chi^2 = 1338.7$ (flux uncertainties based on photon noise only).
\item We may also compare the $\chi^2$ of the entire lightcurve fit (using the model described in \S4.3).  In this case, we must scale the $\chi^2$ for a fair comparison since the linear fit uses fewer points due to the truncation procedure.  Comparing the reduced $\chi^2$ between the two corrections we find lower values for the linear fit again- 1.045 vs 1.014 or 1.031 vs 1.000, depending whether we additionally correct for pixel phase\footnote{Reduced $\chi^2$ values have been rescaled so that lowest value is equal to unity}.
\item We use the fitted transit duration, $T$, defined by Carter et al. (2009) which was shown to be highly robust and non-degenerate.  $T$ is expected to be independent of wavelength as the only possible parameter which could vary is $R_*$ which is not expected to exhibit significant changes between different wavelengths.  We therefore refit the Brown et al. (2001) HST lightcurve of HD 209458b, taken in the visible, with the same model used here.  We find a transit duration of $9525_{-14}^{+16}$ seconds.  In comparison, correcting the second epoch  at 5.8 $\mu$m with a linear fit yields $T_{lin} = 9518_{-55}^{+50}$ seconds and with a logarithmic fit $T_{log} = 9546_{-55}^{+51}$ seconds.
\item In Fig 4., the local gradients, as taken in 150-point bins, is compared to that expected from the logarithmic fit of the data.  There is a very strong discrepancy between the data and the model before 2000 seconds.
\end{enumerate}

Thus, we find employing a logarithmic fit, with two additional free parameters, cannot be shown to offer any kind of improvement over the linear fit.
The gradient analysis presented above shows that the logarithmic parametrisation is not adapted. Moreover, it is disfavored by $\Delta \chi^2=20$. Since every single test performed has supported the truncation and linear trend correction, this method will be adopted at the preferred corrective procedure in our subsequent analysis. 

%For the first and second epoch, the centroid of the PSF is in the pixels (124, 128) and (124, 129) respectively.

\subsection{Correcting the ramp at 8 $\mu$m}

The ramp effect at 8$\mu$m is well documented and so too is the methodology for correcting this phenomenon (Agol et al., 2008 and references therein).  Unlike the 5.8$\mu$m data, there are no known discontinuities in the time series and thus the correction may be achieved using a smooth analytic function.  We fit a time trend to the out-of-transit data of the form $a +b t + c \log (t-t_0 ) + d (\log(t-t_0 ))^2$ where $t_0$ is chosen to be 30 seconds before the observations begin to prevent the function exploding at $t=0$.

\begin{figure*}
\begin{center}
\includegraphics[angle=0,width=9. cm,]{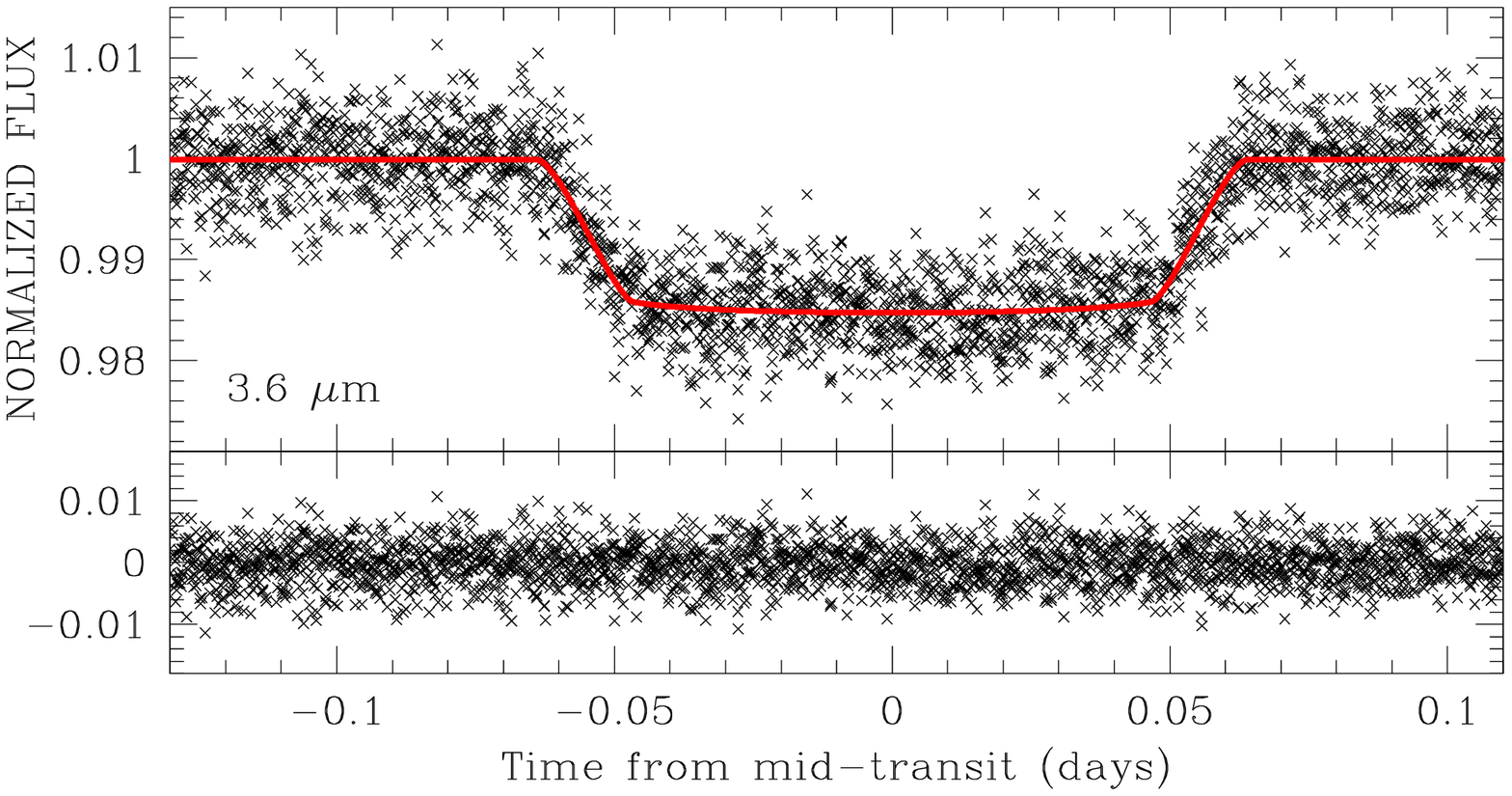}\includegraphics[angle=0,width=9. cm,]{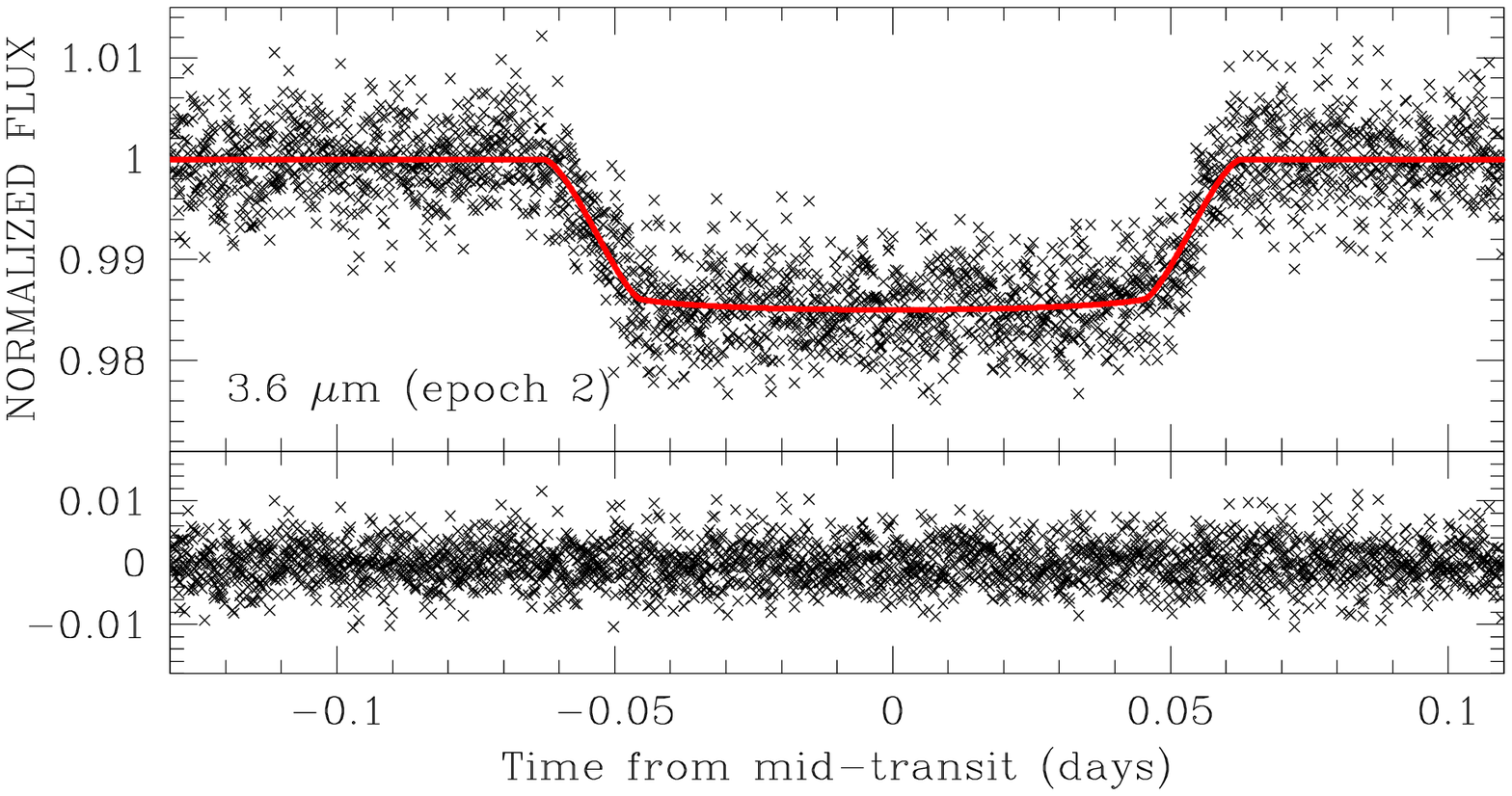}
\includegraphics[angle=0,width=9. cm,]{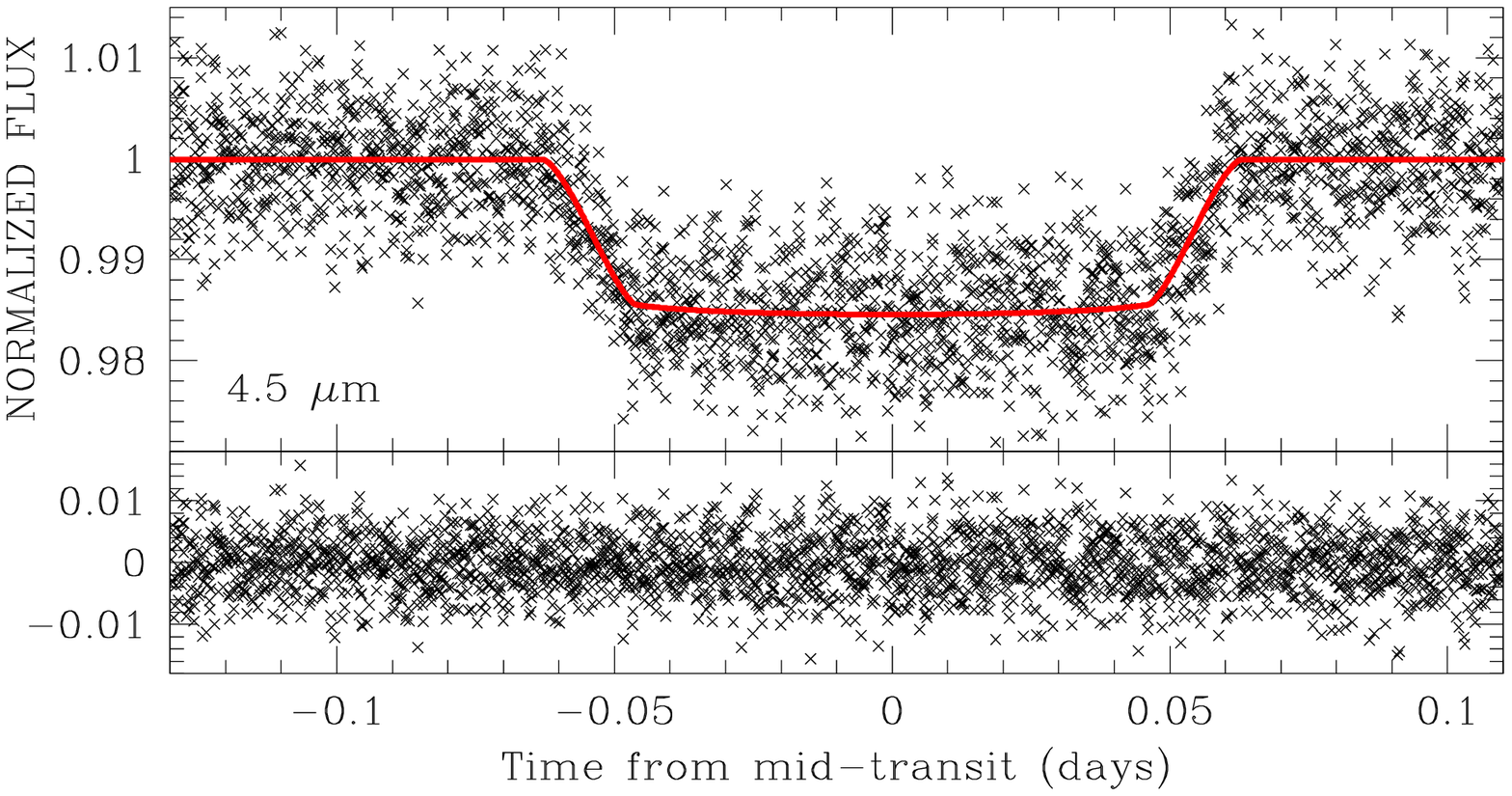}\includegraphics[angle=0,width=9. cm]{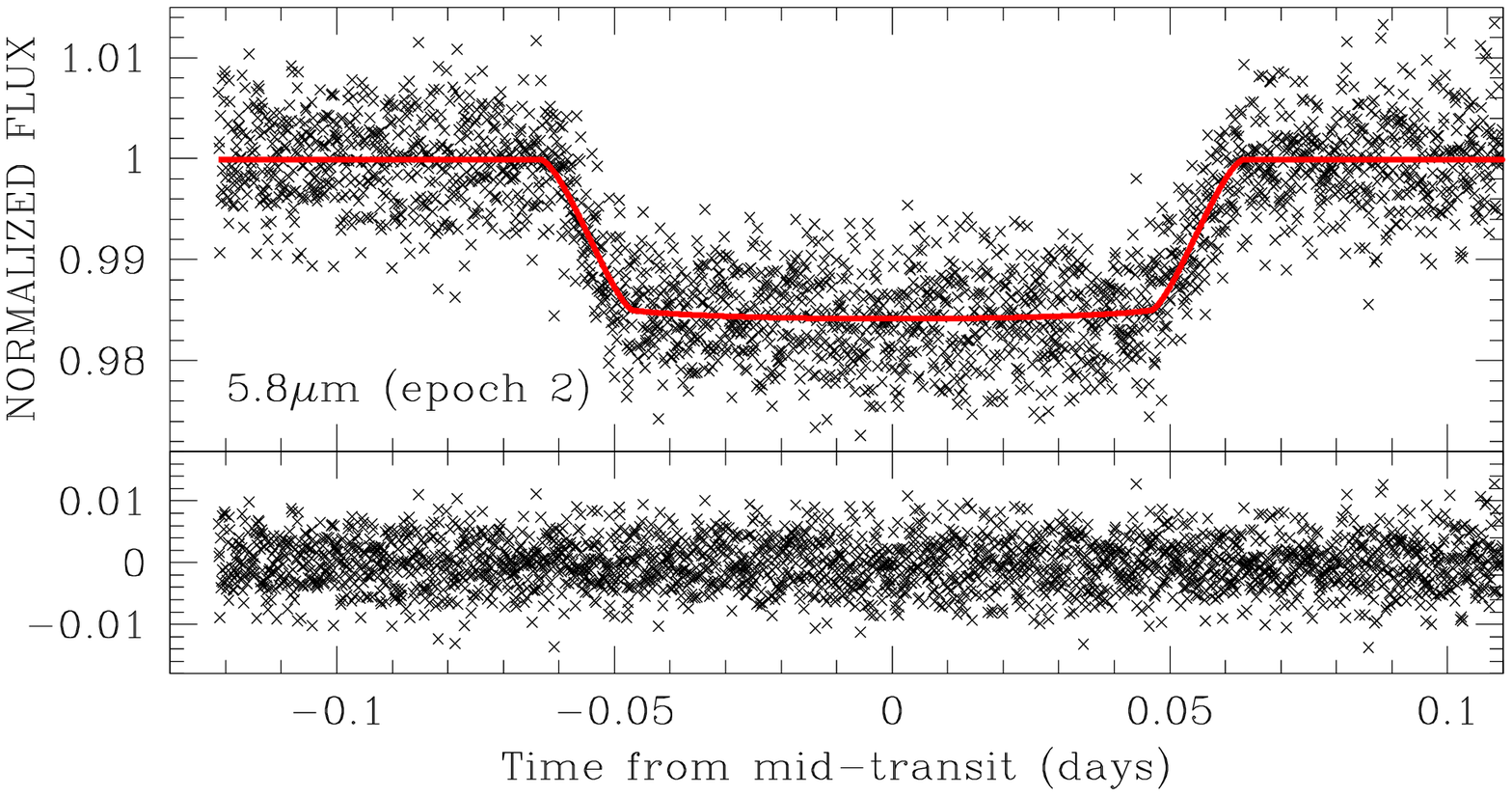}
\includegraphics[angle=0,width=9. cm]{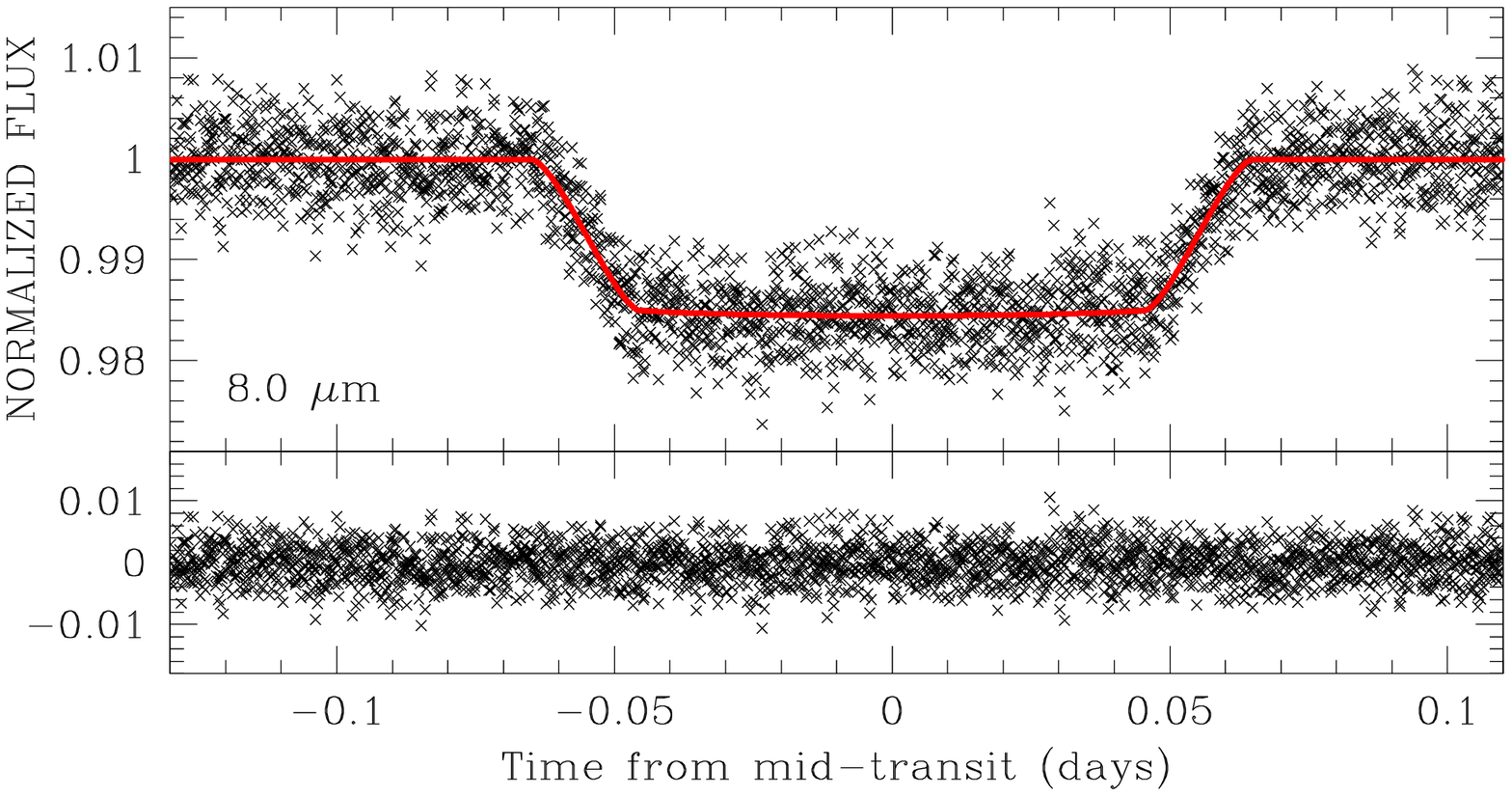}
\caption{\emph{Final light curves, best fit model and residuals at  3.6 $\mu$m (epoch 1 and 2), 
4.5 $\mu$m, 5.8 $\mu$m (epoch 2) and 8 $\mu$m. In the residuals
subpannel we will overplot the 50-point median-stack smoothing of the residuals. 
}} \label{fig:figfinal}
\end{center}
\end{figure*}

\section{Fitting the transit light curves}

Among the 6 transit light curves, we have four of high quality with
well understood and corrected systematic effects : the first epoch at
3.6 $\mu$m, 4.5 $\mu$m, the second epoch at 5.8 $\mu$m and 8 $\mu$m.
The second epoch at 3.6 $\mu$m and the first epoch at 5.8 $\mu$m 
will be treated separately.  

\subsection{Limb darkening}

Accurate limb darkening coefficients were calculated for each of the
four IRAC bands. We adopted the following stellar properties:
$T_{\rm eff}= 6100$ K, $\log g=4.38$, and $[Fe/H]=0$. We employed the 
Kurucz (2006) atmosphere model database providing intensities at 17 
emergent angles, which we interpolated linearly at the adopted $T_{\rm 
eff}$ and $\log g$ values. The passband-convolved intensities at each of 
the emergent angles were calculated following the procedure in Claret 
(2000). To compute the coefficients we considered the following expression:
\[\frac{I(\mu)}{I(1)} = 1-\sum_{k=1}^{4} c_k (1-\mu^{k/2}), \]
where I is the intensity, $\mu$ is the cosine of the emergent angle, and 
$c_k$ are the coefficients. The final coefficients resulted from a least 
squares singular value decomposition fit to 11 of the 17 available 
emergent angles. The reason to eliminate 6 of the angles is avoiding 
excessive weight on the stellar limb by using a uniform sampling (10 
$\mu$ values from 0.1 to 1, plus $\mu=0.05$), as suggested by 
D\'{\i}az-Cordov\'es et al. (1995). The coefficients are given in Table 
\ref {table:ld}.

\subsection{Markov Chain Monte Carlo fit to the data}
We adopt the physical model of a transit light curve through the
expressions of Mandel \& Agol (2002) and orbital eccentricity using
the equations of Kipping (2008).  We sampled the parameter space with
Markov Chain Monte Carlo codes (Doran \& Muller 2004) originally developed for microlensing 
(Dong et al., 2008; Batista et al., 2009) and adapted  to fit transit data. We first made an
independent fit for 3.6 $\mu m$ (epoch 1 and 2), 4.5 $\mu m$ , 5.8 $\mu m$  (epoch 2) and 8 $\mu m$.
We adopted a fixed value of period to be $P=3.524749$ days (Knutson et al. 2007). For each channel, 5
parameters are fitted, namely the out-of-transit baseline, the orbital inclination $i$, the ratio between the
orbital semi-major axis and the stellar radius $a/R*$, the ratio of
radii, $k$, and the mid-time transit $t_c$.  We also permit the orbital eccentricity $e$ and the position of periastron $\omega$ to move in a restricted range, corresponding to the best-fit values derived by Winn
et al. (2005) including their error-bars. The \rjbch{five other} parameters are
free.  The error-bars of the data have been rescaled to make the
$\chi^2$ per degree of freedom equal to unity.  The results are shown in
Table \ref{table:mcmc}.

As some physical parameters should be the same for all bands, we \rjbch{made a}
simultaneous fit to the best observations, namely 3.6 $\mu m$ (epoch 1), 4.5 $\mu m$ , 
5.8 $\mu m$  (epoch 2) and 8 $\mu m$, in which four parameters
are shared by all channels: $P$, $e$, $i$, $\omega$ and $a/R*$. Three
other parameters, $k$, $t_c$ and the baseline, are fitted independently for each
band and are allowed to move within the range obtained in the
individual fits.  We decided to fit  separately 3.6 $\mu m$ (epoch 2), forcing
the four shared parameters to be equal to the values derived from the
best fit with the four other channels. The results are shown in Table \ref{table:mcmcglob}.

\begin{table}
\caption{\emph{ Limb darkening coefficients.}} % title of Table
\begin{tabular}{l r r r r  } % centered columns (5 columns)
\hline 
 channel           &   c1             &     c2            &    c3            &    c4 \\
\hline 
 (3.6 $\mu$m) &  0.2670569 &   0.1396675 & -0.1900802&    0.064018\\
 (4.5 $\mu$m) &  0.3325055 & -0.1999922 &   0.1858255&  -0.0703259\\
 (5.8 $\mu$m) &  0.3269256 & -0.2715499 &  0.2258883 &  -0.0684003\\
(8 $\mu$m)     &  0.2800222 & -0.2278080 &  0.1451840 &  -0.0273881\\
\hline 				
\end{tabular}
\label{table:ld} % is used to refer this table in the text
\end{table}

\begin{table*}
\caption{\emph{ Markov Chain Monte Carlo fit to individual primary transits observed by IRAC.  
We list all the fitted parameters (see the text for the description), 
and in particular the ratio-of-radii, $k=R_p/R_*$, the orbital semi-major axis divided by the stellar radius, $a/R*$, 
the orbital inclination, $i$ and the mid-transit time, $t_c$. }} % title of Table
\begin{tabular}{l l l l ll l l} % centered columns (6 columns)
\hline 
	& 3.6 $\mu$m (epoch 1 ) & 3.6 $\mu$m (epoch 2) & 4.5 $\mu$m & 5.8 $\mu$m (epoch 2) & 8 $\mu$m \\
\hline 	
%$e$           &   $0.0146 \pm 0.0002$ &  $0.0152 \pm  0.0002$    &  $0.0147 \pm 0.0002$     &  $0.0147 \pm  0.0002$  \\		
%$\omega$ &   $84.00 \pm  0.01$  &  $84.01 \pm  0.01$	 &  $84.02 \pm  0.01$ 	& $84.01 \pm  0.01$ \\
%Nb data     & 2858 & 2854  & 2823 & 2511 & 2815 \\
%$\chi^2$   & 2883 & 2828  & 2859 & 2520	&  2778\\
%Rescaling  &  0.92 & 0.94   &0.94& 0.76  &0.46\\
%\hline 				
$ i (deg)$	 & $ 87.00\pm  0.11$ & $ 86.67\pm  0.15$  & $ 86.87 \pm  0.10$	 & $86.84 \pm 0.11$	& $86.37 \pm  0.13$ \\
$A/R_*$    &  $8.89 \pm 0.06$   &  $8.84 \pm 0.10$  & $8.91 \pm  0.05$     & $ 8.84\pm  0.07$	 &  $8.49 \pm  0.08$\\
\hline 
$k=R_p/R_*$  	& $0.120835 \pm  0.00054$ & $0.120387  \pm  0.00053$ &$ 0.1218\pm 0.00072 $  & $0.1244 \pm  0.00059$ & $0.1240 \pm  0.00046$\\
$k^2=(R_p/R_*)^2$	& $1.460 \pm 0.013 \%$  &$ 1.449\pm 0.013 \%$ &  $1.4835 \pm 0.017 \% $& $1.547 \pm 0.015 \% $  & $1.538 \pm 0.011 \% $\\

\hline 
\end{tabular}
\label{table:mcmc} % is used to refer this table in the text
\end{table*}

\begin{table*}
\caption{\emph{ Markov Chain Monte Carlo fit to IRAC data. The first column shows a join fit to the best four band observations,
namely 3.6 $\mu$m (epoch 1) ,  4.5 $\mu$m ,  5.8 $\mu$m (epoch 2) and 8 $\mu$m. Then we impose the parameters $i$, $A/R_*$, $e$, $\omega$ and fit the ratio of the radii  $k$  and the mid transit time $t_c$ for the second epoch at 3.6 $\mu$m and the first epoch at 5.8 $\mu$m .}} % title of Table
\begin{tabular}{l l l l l} % centered columns (6 columns)
\hline 
	&3.6 $\mu$m (epoch 1 )  + 4.5 $\mu$m 	                &  (3.6 $\mu$m) epoch 2 &  (5.8 $\mu$m) epoch 1\\
        & + 5.8 $\mu$m (epoch 2) + 8 $\mu$m  & & \\
\hline 	
%Nb data     &  2888 / 2794 / 2520 / 2783 &  & \\
%$\chi^2$   &  2885 / 2849 / 2556 / 2822 & 2806 & \\
%\hline 
$ i (deg)$	 & $86.76 \pm 0.10$   & \\
$a/R_*$    &  $8.77 \pm  0.07$     & \\
\hline 
$k=(R_p/R_*)$  & & \\
(3.6  $\mu$m)	& $0.121215 \pm 0.00054$  & $0.120343 \pm  0.00053$ \\
(4.5  $\mu$m)	& $0.121568 \pm 0.00072$  &  \\        
(5.8  $\mu$m)	& $0.1244     \pm 0.00059$  & &  $0.1246 \pm  0.00095$\\	
(8     $\mu$m)	& $0.12390     \pm 0.00046$  &  \\	
\hline 
$k^2=(R_p/R_*)^2$ & & \\
(3.6  $\mu$m)	& $ 1.469 \pm 0.013 \%$  & $ 1.448 \pm  0.013 \%$ \\
(4.5  $\mu$m)	& $ 1.478 \pm 0.017 \%$  &  \\        
(5.8  $\mu$m) & $  1.549 \pm 0.015 \%$  & &  $  1.552 \pm 0.032 \%$ \\	
(8.0  $\mu$m)	& $  1.535 \pm 0.011 \%$  &  \\	
\hline 
\end{tabular}
\label{table:mcmcglob} % is used to refer this table in the text
\end{table*}

\subsection{Prayer-bead Monte Carlo  fit to the data}

We also fitted all the transit data with the code used by Fossey et al. (2009), incorporating the effects of non-linear limb darkening through the expressions of  Mandel \& Agol (2002) and orbital eccentricity using the equations of Kipping (2008). We fixed the orbital eccentricity, $e$, and position of periastron, $\varpi$, to the best-fit values derived by Winn et al. (2005), and adjusted $k$, $a/R*$, $b$, and $t_c$ to find a minimum in $\chi^2$. Although the parameters $a/R*$ and $b$ show a degree of covariance, Carter et al. (2008) have shown that the transit duration, $T$, ratio-of-radii, $k$, and mid-transit time, $t_c$, are non-degenerate parameters; these parameters are also less affected by systematic errors in orbital eccentricity and thus can be taken to be more reliably constrained than $a/R*$, $b$, or the inclination, $i$.

We use the genetic algorithm {\sc pikaia} (see Metcalfe \& Charbonneau 2003) to find an initial, approximate
solution, which is used as the starting point for a $\chi^2$-minimisation using the downhill-simplex {\sc amoeba} algorithm 
(Press et al. 1992).  The initial best-fit parameters from {\sc amoeba} are randomly perturbed
by up to 40\% of their value and refitted in 100 trials to check the robustness of the best-fit solution.

To obtain the final parameter uncertainties, we employ a `prayer-bead' Monte Carlo simulation of the unbinned data, as used by Gillon et al. (2007). Here, the set of residuals from the best-fit solution is shifted by one data point and added to the best-fit transit model to generate a new data set, with the  residual at the end of the data series wrapping around to the beginning. The new data set is refitted, and the process repeated until the set of residuals has been cycled through the entire data series. This procedure has the advantage of preserving the structure of any residual  correlated noise within the light curve in each simulation. For the unbinned data we then have typically 2500--3000 samples from which the parameter uncertainties may be estimated, which we take to be the values comprising 68.3\% of the sample about the median of each parameter distribution. The median and uncertainties are compared to the fitted value in each case, to check the robustness of the simulations and to assign upper and lower limits on the parameters. In all cases, we found the difference between  the median and the best-fit parameter was insignificant.  

Table 5 lists the fitted depth, ratio of radii, $k$, transit duration, $T$, orbital inclination, $i$, and $a/R*$ from this fitting procedure, for each of the transits.

\begin{table*}
\caption{\emph{Best-fit transit depths, ratio of radii $k$, duration $T$, orbital semi-major axis divided by 
the stellar radius $a/R_*$, inclination $i$, mid transit time $t_c$  found using the prayer-bead Monte Carlo 
fit method described in \S4.1}} % title of Table
\begin{tabular}{l c c c l l} % centered columns (7 columns)
\hline 
band/$\mu$m &$k^2=(R_p/R_*)^2$ & $k=R_p/R_*$ & $T$/seconds & $a/R_*$ & $i$ \\
\hline 
3.6 $\mu$m (epoch 1) & $1.462_{- 0.012}^{+0.011}$ & $0.12089_{-0.00048}^{+0.00045}$ &  $9581_{-47}^{+57}$ & $8.88 \pm 0.02$ & $86.99 \pm 0.18$ \\
4.5 $\mu$m & $1.482_{-0.014}^{+0.014}$ & $0.12174_{-0.00056}^{+0.00056} $ &  $9437_{-51}^{+60}$ & $9.09 \pm 0.01$ & $87.15  \pm 0.10$ \\
8.0 $\mu$m & $1.538_{-0.011}^{+0.011}$ & $0.12403_{-0.00045}^{+0.00043}$  &  $9580_{-49}^{+58}$ & $8.50 \pm 0.02$ & $86.32 \pm 0.20$ \\
3.6 $\mu$m(epoch 2) & $1.449_{-0.010}^{+0.010}$ & $0.12038_{-0.00043}^{+0.00043}$ &  $9358_{-48}^{+57}$  &  $8.86 \pm 0.02$   &$86.70  \pm 0.13$ \\
5.8 $\mu$m(epoch 2) & $1.542_{-0.0096}^{+0.0099}$ & $0.12416_{-0.00039}^{+0.00040}$ &  $9517_{-54}^{+50}$ &  $9.13  \pm 0.01$ &$87.22 \pm  0.12$ \\
\hline 
\end{tabular}
\label{table:david} % is used to refer this table in the text
\end{table*}

\begin{figure}
\begin{center}
\includegraphics[angle=0,width=9. cm,]{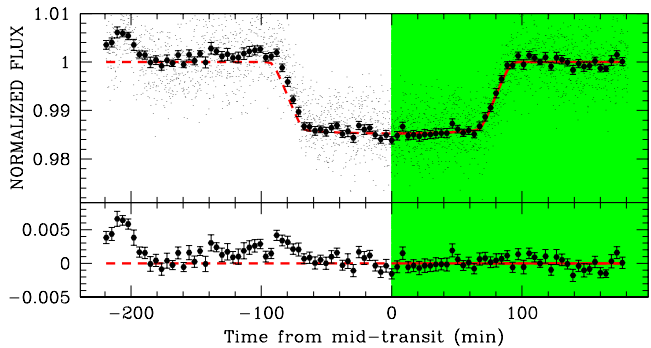}
\caption{\emph{The uncorrected unbinned and binned (30-points) data from first epoch at 5.8  $\mu$m and the underlined model computed for the second epoch (corrected for systematics). The lower pannel shows the residuals of the binned data (unbinned data ommitted for clarity).
The shaded area is marking the second half of the transit and the post transit 
observations used in the fit. }} \label{fig:ch7}
\end{center}
\end{figure}

\subsection{The case of the epoch 1 of 5.8 $\mu$m }

The first epoch of 5.8$\mu$m requires special consideration due to the extremely pronounced nature of the systematic errors for this data set. In \S3.2 we demonstrated that the corrective procedure most consistent with the observational evidence is to the truncate the initial data exhibitting discontinous behaviour and then perform a linear correction through the remaining baseline data.  Since the systematic effect is so pronounced here, we took a conservative approach by assuming the systematic may persist up the moment of mid-transit. We therefore exclude all data before the mid-transit and adopt the physical parameters ($i$, $a/R_*$) derived from the global MCMC fit and reported in Table 4, and fitted the baseline, the ratio of radii $k$ and the mid transit time $t_c$.

In Fig.  \ref {fig:ch7} we compare epoch 1 data and the model fitted on the data 
from the mid-transit. Inspection of the residuals after the mid-transit indicates a good fit to the data. 
The data from the first half of the observations show the uncorrected systematic trends at 
work. It is clear that they are of different nature from the one of epoch 2. We add the measured ratio of radii
and depth in the last column of Table 4, and last row of table 5. 
The results we therefore obtain from second epoch is consistent with the first epoch. 

As a final check of the procedure, we decided to treat also the second epoch at 5.8 $\mu$m  the same way. 
We take the uncorrected data, exclude the first half of the data up to the mid-transit, and fit the light curve.
We report the measured depth by this procedure to be $1.540 \pm 0.029 \%$. It is perfectly compatible 
with our complete fits reported in Tables 3, 4 and 5.

\subsection{Sanity check: Grid calculations}

As a \rjbch{check on our methodology} we \rjbch{imposed} the physical
parameters derived from Knutson et al. (2008) in the approximation of
circular orbit, and \rjbch{fitted} for the baselines, ratio of radii,
$k$, and \rjbch{mid-time} transit, $t_c$,  using a simple $\chi^2$ minimisation.
This gradient based method explores the local minimum around the physical solution found by Knutson et al., (2008). By comparing the results to the ones obtained with the Monte Carlo methods, we 
found that the transit depths agree well within the error bars. 
By contrast the other parameters (inclination, $a/R_*$) are highly degenerate.

\subsection{Influence of spots}
An effect to consider when comparing transit depth at different 
wavelengths is the influence of stellar surface inhomogeneities, i.e, 
star spots. Depending on the spot distribution, the occulted stellar 
area during the transit can be brighter or dimmer than the average 
photosphere. In the case of HD 189733, a moderately active star with 
visible photometric variations of $\sim$3 \% (peak to peak), the 
differential effect in the IRAC 3.6 and 5.8 $\mu$m bands was evaluated 
by Beaulieu et al. (2008) to be below 0.01\%. HD 209458 is a 
chromospherically inactive star with an estimated age close to that of 
the Sun (e.g., Mazeh et al. 2000; Cody \& Sasselov 2002; Torres et al. 
2008). It is thus reasonable to assume a level of photometric variations 
similar to that of the Sun, i.e., 0.2--0.3\% peak to peak (Fr{\"o}hlich 
\& Lean 2004). Scaling the calculations carried out for HD 189733, the 
expected differential effect of star spots on the IRAC bands for HD 
209458 is likely to be 10-20 times smaller, and therefore well below 
0.001\%. Our calculations show that star spots have negligible influence 
when compared with our measurement uncertainties ($\sim$0.011-0.017\%); 
see Tables \ref{table:mcmc},\ref{table:mcmcglob} and \ref{table:david}.

\subsection{Comments about different epochs at 3.6 and 5.8 $\mu$m  }
We asked for two epochs for HD~209458b at 3.6 and 5.8 $\mu$m with the
prime intention of demonstrating the possibility of \rjbch{co-adding} multiple
epoch observations, and/or to be able to check for the variability in the
system. The two epochs are separated by 7 months, and the observing
setups are identical.

\rjbch{Firstly}, at 3.6 $\mu$m the data are affected by systematic
trends of the same nature due to the pixel scale effect.  We notice a
factor 3 in the amplitude of the systematic trends between the two
epochs. We measure the two transit depth to be $ 1.469 \pm 0.013 \%$  
and  $ 1.448 \pm  0.013 \%$ respectively. The results are compatible between 
the two channels.

Secondly, at 5.8 $\mu$m the situation is more complex.  
The second epoch showed the expected \rjbch{behaviour}, and we have been able to
correct for systematics, and to fit it. For the first epoch, we choosed to discard the 
first half of the data, and fit the uncorrected remaining data set.
We measure the two transit depth to be   $1.552 \pm 0.032 \%$ 
and $  1.549 \pm 0.015 \%$ respectively. The results are compatible between  
the two channels.

Even when centering on the same pixels of the detector, 
observing the same target several months apart, different systematics are at work.
It is clear that the different data sets should be analysed for systematics and 
then corrected individually. Then, multiple epoch can be compared and/or added. 

\subsection{Results}

We have chosen three approaches to fit the data, i.e., grid calculations with ephemeris from Knutson et al., 2008, Markov Chain Monte Carlo and prayer-bead Monte Carlo. We obtain extremely similar results concerning the transit depth for the different wavelengths with the three techniques. The final results are listed in table \ref{table:mcmcglob}.  As reported by Carter al. (2008), there exists a degeneracy between the fitted orbital inclination $i$ and $a/R_*$; whereas the ratio of radii $k$ and the transit duration, $T$, are far more robust quantities.  As a result of this robustness, we are able to use the fitted transit duration values as a test of whether the lightcurves appear physical or not.

From our six fitted lightcurves, the duration of 5.8$\mu$m, epoch 1, cannot be used because only half the transit is fitted and so the fitted duration is dependent on priors.   The other five lightcurves produce durations consistent with an average duration of $T = 9508 \pm 17$ seconds with a $\chi_{red}^2 = 2.6$ suggesting an outlier.  Removing the 3.6$\mu$m, epoch 2, measurement to leave us just the 4 preferred observations we find $T = 9538 \pm 14$ seconds with $\chi_{red}^2 = 1.1$.  Since 3.6$\mu$m, epoch 2, produces an outlier duration and is also known to exhibit by far the strongest pixel phase effect out of all of the observed lightcurves, we give it zero weighting in the later spectral analysis.

We find that our average durations are consistent with the duration we find when refitting the Brown et al. (2001) HST lightcurve of $T = 9525_{-14}^{+16}$ seconds.  Consequently, we conclude that our results support a solution consistent with the Brown et al. (2001) observations and thus we may be confident that the systematic corrections have been successful.

\subsection{Comparison with HD~189733b data}

\rjbch{Beaulieu et al. (2008) gives an accurate description of the method adopted} to analyse the two IRAC channels at 3.6 and 5.8 $\mu$m
in the case of the hot-Jupiter HD~189733b.  \rjbch{The software BLUE (Alard, 2010), was used} to fit the PSF, as several stars were in the field and could be used as calibrators. One of the \rjbch{capabilities} of BLUE is to provide optimised centroid estimates, and provide an accurate modelling of the PSF. In the case of HD~209458b one star only was present, so we had to adopt a different strategy, using the SExtractor programme (Bertin and Arnouts, 1996). The extracted light curves at 3.6 and 5.8 $\mu$m were corrected \rjbch{in a similar manner to that detailed in} Beaulieu et al. (2008).  In particular the 3.6 $\mu$m observations for the two planets \rjbch{show} moderate or strong pixel-phase effects, that \rjbch{can be corrected} for.
\begin{figure}
\begin{center}
\includegraphics[angle=0,width=8.5 cm,]{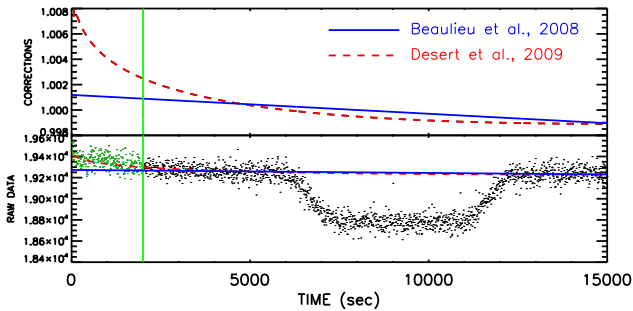}
\caption{The lower pannel shows the reprocessed HD 189733b data at 5.8  $\mu$m  overplotted with the logarithmic correction
from Desert et al. (2009) and the linear correction (Beaulieu et al. 2008). The vertical line indicates 2000 sec. We provided in the
text evidences to reject the first 2000 sec. This figure shows how the logarithmic correction is overcorrecting in the transit.} \label{fig:189bad}
\end{center}
\end{figure}

5.8 $\mu$m observations ostensibly represents the greatest challenge for correcting systematic errors as the behaviour is somewhat less understood than the 8.0 $\mu$m  ramp and the pixel phase effects at 3.6 $\mu$m and 4.5 $\mu$m.  This situation is exacerbated by the observation of a discontinuity in the photometry in two separate transit observations.  At 5.8 $\mu$m \rjbch{there are no} significant pixel-phase effects, but a linear drift with time, after the first 2800 s in the case of HD~209458b and 1800 s in the case of HD~189733b. For both planets we disregard the first 2800/1800 seconds and then simply apply a linear correction to the data after this point. 

This is the main discrepancy between the Beaulieu et al. (2008) reduction and the one adopted by D\'esert et al. (2009). We do not discuss here previous results or methods adopted by the same team, incorporated in Ehrenreich et al. (2007), in part because we have already explained the reasons of their inadequacy in the Beaulieu et al. (2008), but most importantly as clearly abandoned by the authors  themselves in the new version of the analysis of the same data provided by D\'esert et al. (2009). D\'esert et al. (2009) applied a logarithmic time-correlated detrending to this channel of the same form of 8.0 $\mu$m observations, but of opposite sign i.e. an `anti-ramp'.  In contrast, we applied a truncation of the first 2800 seconds followed by a linear time-trend de-correlation.

The question as to which method is the correct one is naturally a topic of debate within the community but we believe we have produced here an in-depth analysis of each type of correction.  In \S3.2 we treated the 5.8$\mu$m data with both methods and compared the resultant lightcurves.  Several tests suggest that the truncation + linear detrending produces a more physical transit signal and an improved overall fit.  In this work, we acknowledge that there currently exists no widely-accepted physical explanation for the channel 3 systematic effects and thus the preference between the linear and logarithmic model must be made primarily on the basis of the lightcurve information.  On this basis, we cannot justify employing the logarithmic model over the method adopted here, given the range of evidences compiled.

In our case, we find that adopting the logarithmic fit to our HD 209458 5.8$\mu$m data would underestimate the transit depth by 0.035 \%, generating a systematic error of $\sim 2.3 \sigma$.  We also estimate that the incorrect use of logarithmic correction leads to an under estimate of the transit depth of HD 189733 by 0.047\% , generating a systematic error of $\sim 2.9 \sigma$. This accounts for the discrepancy between the studies of D\'esert et al. 2009 compared to Beaulieu et al. 2008.
 However, for case of 3.6 $\mu$m, both teams agree upon the corrective procedure, and so we should expect very similar results. Indeed, for the HD 189733 3.6$\mu$m photometry, the values and error bars estimated by Beaulieu (2008) and D\'esert (2009) are in excellent agreement, as shown in Table \ref{table:comp3}. 

\begin{table}
\caption{\emph{Comparison of values of transit depth for HD 189733b at 3.6 and 5.8  $\mu$m 
by Beaulieu et al., (2008), Ehrenreich et al., (2007), and D\' esert et al., (2009). }} % title of Table

\begin{tabular}{c c c c } % centered columns (4 columns)
\hline \hline
IRAC                    &  Beaulieu 2008         & Ehrenreich 2007             &D \' esert 2009 \\
\hline
3.6 $\mu$m & $2.383 \pm 0.014 \%$ & $2.434  \pm 0.026\%$     & $2.387  \pm 0.0093 \%$ \\
%4.5 $\mu$m &                                       &                                            & $2.420  \pm 0.0093\%$\\
5.8 $\mu$m & $2.457 \pm 0.017 \%$  & $2.375  \pm  0.04 \%$   & $2.393 \pm 0.016 \%$\\
%8.0 $\mu$m &  $2.380  \pm 0.020 \%$  &                                          & $2.383 \pm 0.013 \%$\\
\hline \hline
\end{tabular}
\label{table:comp3} % is used to refer this table in the text
\end{table}

\section{Data interpretation }

To interpret the data, we used the radiative transfer models
described in Tinetti et al. (2007a,b) and
consider haze opacity, including its treatment in Griffith, Yelle and
Marley (1998).

\begin{figure}
\begin{center}
\includegraphics[angle=0,width=9 cm,]{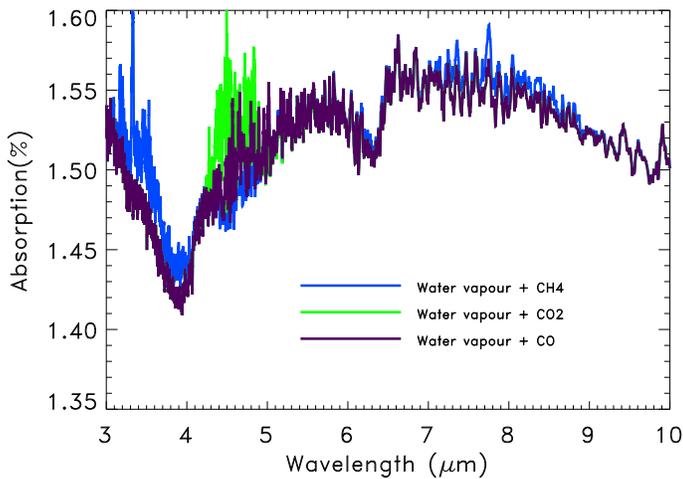}
\caption{\emph{Simulated middle Infrared spectra of the transiting Hot Jupiter HD 209458b in the
wavelength range 3-10 $\mu$m. Water absorption is responsible for the  main pattern of the spectra.
The additional presence of methane, CO and $CO_2$ are simulated in the blue, violet and 
green spectra respectively.}} \label{fig:figcompo}
\end{center}
\end{figure}

Our analysis includes the effects of water, methane,
carbon dioxide, carbon monoxide, pressure-induced absorption of $H_2-H_2$.
We do not consider the presence of particulates, because there is no
indication of particles large enough ($\sim 3 \mu$m ) to affect
the planet's middle-infrared spectrum.
The effects of water absorption are quantified with the BT2 water line list
(Barber et al., 2006), which characterises water absorption at the range
of temperatures probed in HD 209458b.
Methane \rjbch{was} simulated by using a
combination of HITRAN 2004 (\rjbch{Rothman} et al., 2005) and PNNL
data-lists.  Carbon monoxide absorption coefficients \rjbch{were}
estimated with HITEMP (\rjbch{Rothman} et al. 1995) \rjbch{whilst for
carbon dioxide we employed a} combination of HITEMP and CDSD-1000
(Carbon Dioxide Spectroscopic Databank version for high temperature
applications; Tashkun and Perevalov, 2008). \rjbch{The continuum
was computed using $H_2-H_2$ absorption data} (Borysow et al., 2001). In
fig. \ref{fig:figcompo}, we show the contribution of the different molecules combined to
water.

The  3.6 $\mu$m  (and to a lesser degree the one at 8 $\mu$m) IRAC channel
measurement can be affected by the presence of methane. By contrast, CO$_2$ and CO may 
contribute in the passband at 4.5  $\mu$m.

\begin{figure}
\begin{center}
\includegraphics[angle=0,width=9. cm,]{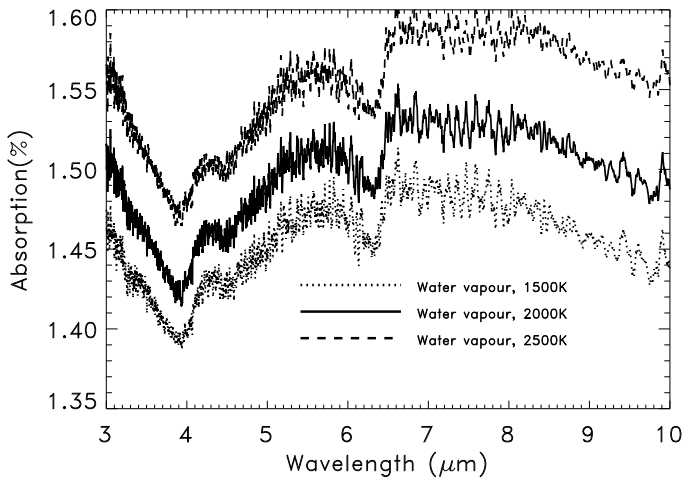}
\caption{\emph{Modeled spectral absorptions of $H_2O$ in the atmosphere of HD 209458b for 1500, 2000 and 2500 K. }} \label{fig:figtemp}
\end{center}
\end{figure}

\begin{figure*}
\begin{center}
\includegraphics[angle=0,width=18. cm,]{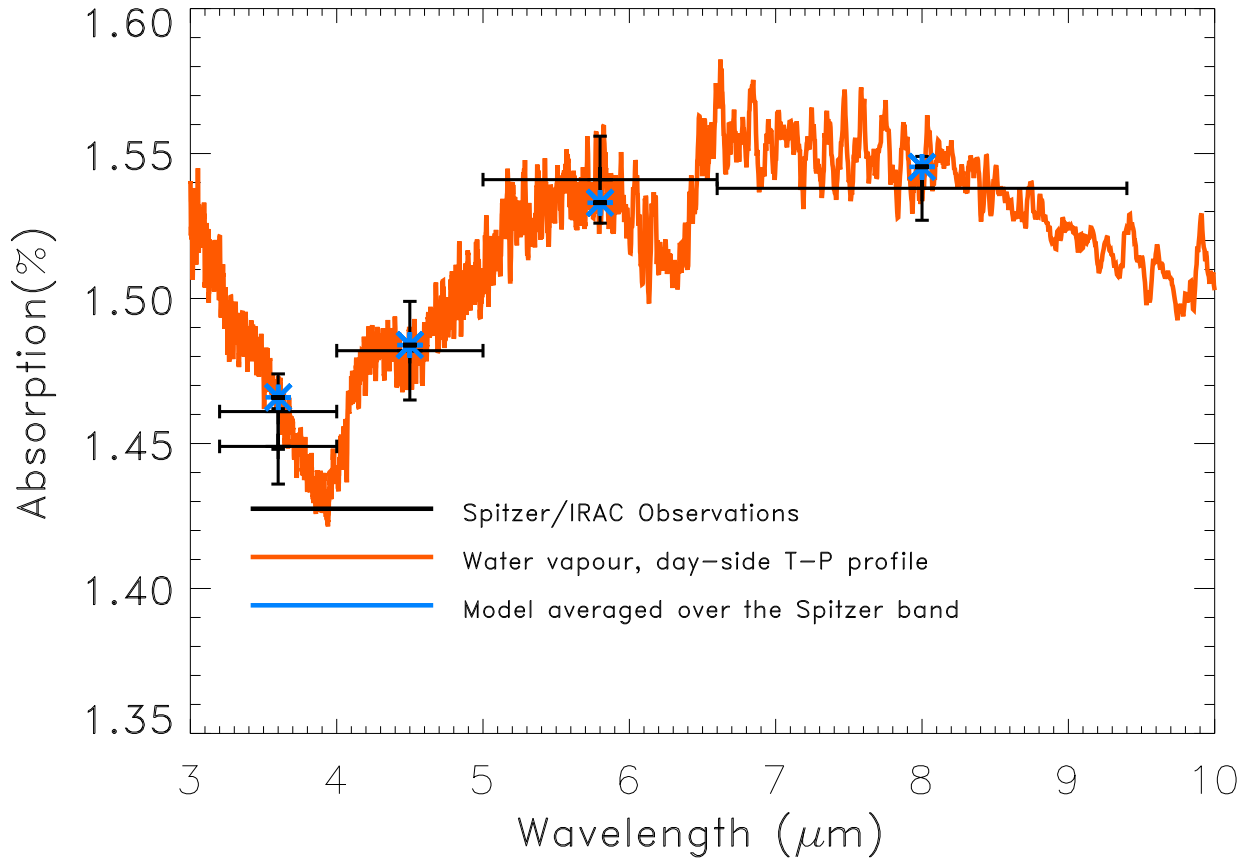}
\caption{\emph{Observations and spectral simulations of the atmosphere of HD~209458b.  Black : Spitzer measurements
where the horizontal bar is the IRAC  bandwidth. Orange spectrum: water vapour and a thermal profile compatible with the day-side spectroscopy and photometry data (Swain et al., accepted; Griffith and Tinetti, in prep.).  
Blue stars : simulated spectrum integrated over the Spitzer bands.  }} \label{fig:figmod}
\end{center}
\end{figure*}

We find absorption by water alone can explain the spectral
characteristics of the photometric measurements, which probe pressure
levels from 1 to 0.001 bars  (fig. \ref{fig:figmod}). The determined water abundance
depends on the assumed temperature profile and
planetary radius. We find that the data can be interpreted,   with a
thermochemical equilibrium water abundance of $4.5 \times 10^{-4}$ (Liang et al.,
2003;2004), assuming  temperature profiles  from Swain
et al., 2009b. However, $\sim 1\%$ difference in the estimate of the
planetary radius, is compatible with
water abundances 10 times smaller or larger, or with an overall change in
the atmospheric temperature of about $\sim$ 500K (see fig. \ref{fig:figtemp}).
Additional primary transit data at different wavelengths are needed to
improve the constraint.

While the contribution of other constituents is not necessary
to interpret the measurements,  mixing ratios of $10^{-7}$,  $10^{-6}$ and
$10^{-4}$ of CO$_2$, CH$_4$ and CO, respectively, are allowed in our
nominal model.

  Spectroscopic data are needed to further investigate
the composition  of this planetary atmosphere.

\section{Conclusion}

We have presented here IRAC photometry data recording the primary
transit of HD~209458b in four infrared bands. We find that the
systematics are very similar to \rjbch{those present in the} data set
obtained for the planet HD~189733b (Beaulieu et al., 2008), and
therefore we adopted similar recipes to correct for them.  We have
performed Markov Chain Monte Carlo and prayer-bead Monte Carlo \rjbch{fits} to the data 
obtaining almost identical results.  \rjbch{Our observations indicate the presence of water
vapour in the atmosphere of HD~209458b, confirming previous detections of this molecule
  with different techniques/instruments. Interestingly, the thermal profiles derived for the day-side  
are compatible with this set of data probing essentially the planetary terminator.
 It is possible that additional
molecules, such as methane, CO and/or CO$_2$ are also present, but the
lack of spectral resolution of our data have prevented these from being
detected. Additional data in transmission at different wavelengths
and/or higher resolution will be required to gain information about these other
molecules.}

\section*{Acknowledgements}
We are very grateful to Tommi Koskinen, Alan Aylward
Steve Miller, Jean-Pierre Maillard and Giusi Micela for their
insightful comments.  G.T. is supported by a Royal Society University
Research Fellowship, D.K. by STFC, R.J.B. by the Leverhulme Trust, 
G.C. is supported by Ateneo Federato della Scienza e della Tecnologia - Universit\'a 
di Roma ''La Sapienza'', Collegio univ. "Don N. Mazza" and LLP-Erasmus Student Placement.
We acknowledge the support by ANR-06-BLAN-0416 and the ``Programme Origine
des Plan\`etes et de la Vie''.  This work is based on
observations made with the \emph{Spitzer Space Telescope}, which is
operated by the Jet Propulsion Laboratory, California Institute of
Technology under a contract with NASA.

\label{lastpage}

\end{document}